\documentclass[11pt]{article}      
\pdfoutput=1
\usepackage{amsthm,amsmath,amssymb,xcolor,sectsty,latexsym,mathtools,hyperref,mathrsfs}
\usepackage{slashed}  
\usepackage{bm}
\usepackage{epsfig}
\usepackage{dcolumn}
\definecolor{vermelho}{cmyk}{0,.88,.77,.40}
\usepackage{cite}
\usepackage{feynmp}
\numberwithin{equation}{section}
\usepackage[textwidth=6.5in,top=1.2in,bottom=1.2in]{geometry}
\usepackage{setspace}
\chapterfont{\color{vermelho}}
\usepackage{simplewick}
\usepackage{hyperref}
\usepackage{float}

\newcommand{\be}{\begin{equation}}
\newcommand{\ee}{\end{equation}}
\newcommand{\beq}{\begin{equation}}
\newcommand{\eeq}{\end{equation}}
\newcommand{\ba}{\begin{eqnarray}}
\newcommand{\ea}{\end{eqnarray}}
\newcommand{\bef}{\begin{figure}}
\newcommand{\eef}{\end{figure}}
\newcommand{\amp}{&\!\!\!}

\def\q{{\bf q}}
\def\x{{\bf x}}
\def\y{{\bf \hat{x}}}
\def\k{{\bf k}}

\newcommand{\bra}[1]{\left\langle #1 \right|}
\newcommand{\ket}[1]{\left|#1\right\rangle}
\newcommand{\braket}[2]{\left\langle {#1} \middle| {#2} \right\rangle}

\newcommand{\dv}{\hat{p}}
\newcommand{\DM}{\text{DM}}
\newcommand{\ch}{\chi}
\newcommand{\om}{\xi}

\newcommand{\potl}{X}

\newcommand{\Q}{\mathcal{Q}}
\newcommand{\Psii}{\Psi_{\mbox{\tiny{ini}}}}
\newcommand{\Psif}{\Psi_{\mbox{\tiny{fin}}}}
\newcommand{\rhoDM}{\rho_{\mbox{\tiny{DM}}}}
\newcommand{\tcoh}{t_{\mbox{\tiny{coh}}}}
\newcommand{\cc}{a}
\newcommand{\ccc}{c}
\newcommand{\Gamd}{\Gamma_{\mbox{\tiny{dec}}}}
\newcommand{\tid}{t_{\mbox{\tiny{dec}}}}
\newcommand{\transferbf}{{\bf p}_{tr}}

\begin{document}

\thispagestyle{empty}
\begin{titlepage}
\nopagebreak

\title{  \begin{center}\bf Gravitational Decoherence of Dark Matter\end{center} }

\vfill
\author{Itamar Allali\footnote{itamar.allali@tufts.edu}, ~ Mark P.~Hertzberg\footnote{mark.hertzberg@tufts.edu}}
\date{ }

\maketitle

\begin{center}
	\vspace{-0.7cm}
	{\it  Institute of Cosmology, Department of Physics and Astronomy}\\
	{\it  Tufts University, Medford, MA 02155, USA}
	\end{center}
\bigskip

\begin{abstract}
Decoherence describes the tendency of quantum sub-systems to dynamically lose their quantum character. This happens when the quantum sub-system of interest interacts and becomes entangled with an environment that is traced out. For ordinary macroscopic systems, electromagnetic and other interactions cause rapid decoherence. However, dark matter (DM) may have the unique possibility of exhibiting naturally prolonged macroscopic quantum properties due to its weak coupling to its environment, particularly if it only interacts gravitationally. In this work, we compute the rate of decoherence for light DM in the galaxy, where a local density has its mass, size, and location in a quantum superposition. The decoherence is via the gravitational interaction of the DM overdensity with its environment, provided by ordinary matter. We focus on relatively robust configurations: DM perturbations that involve an overdensity followed by an underdensity, with no monopole, such that it is only observable at relatively close distances. We use non-relativistic scattering theory with a Newtonian potential generated by the overdensity to determine how a probe particle scatters off of it and thereby becomes entangled. As an application, we consider light scalar DM, including axions. In the galactic halo, we use diffuse hydrogen as the environment, while near the earth, we use air as the environment.  For an overdensity whose size is the typical DM de Broglie wavelength, we find that the decoherence rate in the halo is higher than the present Hubble rate for DM masses $m_a \lesssim 5 \times 10^{-7}$\,eV and in earth based experiments it is higher than the classical field coherence rate for $m_a \lesssim 10^{-6}$\,eV. When spreading of the states occurs, the rates can become much faster, as we quantify. Also, we establish that DM BECs decohere very rapidly and so are very well described by classical field theory.
\end{abstract}
\end{titlepage}

\setcounter{page}{2}

\tableofcontents

\newpage

\section{Introduction}

Quantum mechanics allows for the continual generation of macroscopic superpositions of states. These are sometimes referred to as ``Schr\"{o}dinger-cat" states. However, under ordinary circumstances in the everyday world, one does not typically see such macroscopic superpositions. The reason for this has been well established to be due to (i) entanglement and (ii) course graining, as follows: Entanglement inevitably occurs when particles in the environment interact with the Schr\"{o}dinger-cat state; then, by not tracking the environment carefully and focusing only on the sub-system of interest (i.e., coarse graining one's point of view), the quantum coherence becomes effectively destroyed (for early work, see Refs.~\cite{Zeh1970,Zurek1981,Zurek1982}). This is the well known phenomena of {\em decoherence}. Here, we use the word coherent to refer to the full pure quantum mechanical state which evolves unitarily through the Schr\"{o}dinger equation. Thus, decoherence is the process which converts the sub-system of interest into an effective ``mixed state." In this case, the various observable states that made up the reduced superposition are related only by classical probabilities, and thus truly quantum mechanical phenomena, like interference, are not directly observed.
 
For most familiar macroscopic systems,  decoherence occurs very rapidly through ordinary interactions, such as electromagnetic interactions. Typical macroscopic quantum systems readily interact with their surrounding environment, such as air, radiation, etc., which effectively makes a ``measurement'' on the sub-system of interest. Since the role of the environment is often played by a huge number of surrounding degrees of freedom, the subsequent decoherence is usually extremely fast, often in a tiny fraction of a second.  

Given this situation, one may wonder if there are any interesting systems that may be robust against decoherence and thereby maintain their quantum properties for long times. In this work we shall take our system of interest to be dark matter (DM), which comprises most of the mass in the universe. DM is most likely built out of particles beyond the Standard Model (SM). So far, its direct detection has eluded all current experiments. Therefore, we know that DM (unless it is made out of extremely heavy particles) has at most very weak interactions with known SM particles. In fact, it is entirely possible that DM has no interactions at all with the SM (other than through highly suppressed higher dimension operators), and furthermore, it may have extremely suppressed or no interactions with itself and any other particles (beyond the SM). In this situation, a Schr\"{o}dinger-cat-like state of DM would appear to be entirely robust against decoherence from an environment if there is nothing to appreciably interact with.

The obvious exception is gravity, which couples to any source of energy and momentum and thus is coupled to the DM with the exact same strength $G$ as all particles. As a result, a macroscopic quantum superposition of DM may primarily experience decoherence through gravitational interactions. The reason being that the macroscopic superposition creates a gravitational field in a macroscopic superposition too. (Some authors have promoted ``semi-classical" gravity which assumes the gravitational field arises from the expectation value of the mass distribution $G_{\mu\nu}=8\pi G\langle \hat{T}_{\mu\nu}\rangle$, but this extreme point of view will be ignored here).
But importantly, since gravitation is ordinarily so weak, the rate of decoherence may be so slow that DM may maintain its quantum coherence for very long timescales. There has been much work done on decoherence in the context of gravitation and cosmology; see Refs.~\cite{Bassi:2017szd,Belenchia:2018szb,Asprea:2019dok,Anastopoulos:2013zya,Blencowe:2012mp,Breuer:2008rh,Shariati:2016mty,DeLisle:2019dyw,Orlando:2016pwg,Pang:2016foq,Oniga:2015lro,Bonder:2015hja,Diosi:2015vra,Colin:2014vfa,Hu:2014kia,Pikovski:2013qwa,Polarski:1995jg,Halliwell:1989vw,Kiefer:1998qe,Padmanabhan:1989rm,Kafri:2014zsa,Nelson:2016kjm,Anastopoulos:2014yja,Wang:2006vh,Kok:2003mc,Pikovski:2015wwa,Kiefer:1999gt,Mavromatos:2007hv,Tegmark:2011pi,Anastopoulos:1995ya,Colin:2014cfa,Kiefer:2008ku,Brandenberger:1990bx}. 
The goal of this work is to  explore the rate of decoherence of over/under densities of DM in the galaxy, which have organized into macroscopic superpositions of different mass density distributions. Since gravity itself is a non-linear interaction, it is plausible that such quantum states emerge, as the wave function tends to spread from its initial state. This is especially appreciable whenever the system exhibits some form of chaos (e.g., see Ref.~\cite{Albrecht:2012zp}). 

Our primary motivation is very light bosonic DM, especially axions. Since DM is non-relativistic, such particles have a very large de Broglie wavelength $\lambda=2\pi\hbar/(m_a v_a)$. Plausible values of axion masses, including the QCD axions, string theory axions, and axion-like particles, span anywhere from ultra-light axions $m_a\sim 10^{-21}$\,eV$/c^2$ to upwards of $m_a\sim 10^{-3}$\,eV$/c^2$, or so. This corresponds to de Broglie wavelengths spanning $\sim$\,100\,pc to downwards of $\sim$\,meters. So unlike typical de Broglie wavelengths associated with familiar particles, like electrons or protons, the corresponding scales for light DM can be macroscopically large. Furthermore, DM in the galaxy will virialize, leading to a large spread in velocities $v_a$. This means that regular diffuse DM in the galaxy is expected to have $\mathcal{O}(1)$ fluctuations in mass density on the scale of these large de Broglie wavelengths. Furthermore, we will also discuss more compact DM structures, namely boson stars, which arise in some contexts and are very massive configurations of condensed bosons.

This by itself does not make it a Schr\"{o}dinger-cat-like state. In fact, these states are often thought to be well described by {\em classical field theory}, as they correspond to states of extremely high occupancy numbers (e.g., see\cite{Davidson:2014hfa,Guth:2014hsa}). Nevertheless, the interesting point is that the structures are macroscopically large. So {\em if} they evolve into quantum superpositions of distinct classical field configurations (and as we mentioned above, chaotic systems tend to this readily), then we would have macroscopically large Schr\"{o}dinger-cat states in the galaxy. The observational consequences are unclear, since even some Schr\"{o}dinger-cat states can be mimicked by classical ensemble averaging \cite{Hertzberg:2016tal}. But the very persistence of such states is intrinsically interesting and is our focus here.

We will refer to this configuration as a {\em dark-matter-Schr\"{o}dinger-cat-state} (DMSCS). 
Our primary goal is to determine the timescale for decoherence of a DMSCS from the environment of regular matter. We apply these results both to DM in the galactic halo as well as to DM near the surface of the earth where many ongoing experiments are performed.

Our paper is organized as follows:
In Section \ref{Decoherence}, we present the basic formalism to analyze decoherence.
In Section \ref{Scattering}, we perform a non-relativistic quantum scattering calculation and compute general formulae for the decoherence rate.
In Section \ref{Numerics}, we apply these results to (axion) DM and find quantitative results for the decoherence rate.
In Section \ref{Conclusions}, we discuss our results. Finally, in Appendix \ref{app:overlapcomp}, we provide some additional details.

\section{Decoherence Formalism}\label{Decoherence}

In this work we are interested in quantum systems that are in macroscopic quantum superpositions. We will imagine that the DM organizes into such a state. In fact, significant spreading of the wave function is so ubiquitous that in many cases this is inevitable given that the DM has been in existence for billions of years. We will not delve into the details of the formation of such a state, but we just note that it is plausible that it will emerge in some settings. The question of interest is whether such a Schr\"{o}dinger-cat-like state can continue to persist when there is inevitably an environment of some form that it will interact with. In this section, we will lay out the basic formalism for the study of quantum decoherence from gravitation and in later sections we will apply this to DM and determine the decoherence rates quantitatively. 

\subsection{Entanglement}

We will consider a DMSCS, that is a superposition of two distinct observable macroscopic states. Although we will focus on only two states, the extension to an arbitrary number of distinct states is straightforward. These states will be assumed to be distinct mass distributions. We will denote the initial state ket for this as $\ket{\DM}$. The two observable states will be denoted by $\ket{\DM_1}$ and $\ket{\DM_2}$, and its initial state is the superposition
\beq \ket{\DM}=\ket{\DM_1}+\ket{\DM_2}\eeq

\begin{figure}[t]
\centering
\includegraphics[width=\columnwidth]{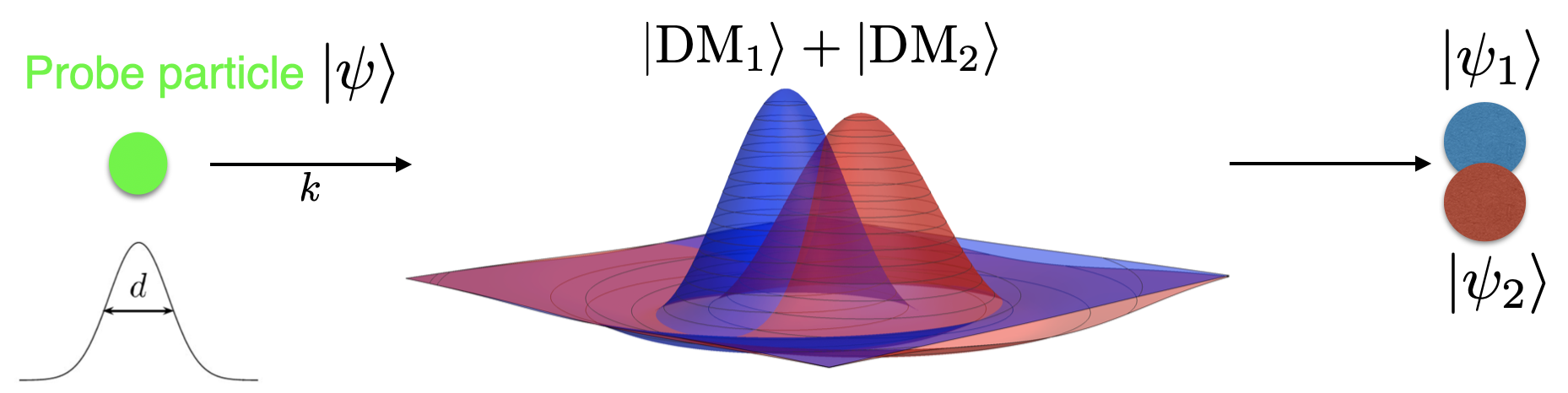}
\caption{A probe particle (wave packet of wavenumber $k$, width $d$) in a quantum state $\ket{\psi}$ approaches a dark-matter-Schr\"{o}dinger-cat-state (DMSCS). The parts of the superposition may be a random perturbation about some background DM density, which takes the form of an overdensity surrounded by an underdensity, whose integrated mass is zero. As a result, the probe particle is not subject to a significant gravitational interaction until close approach. The DMSCS is a quantum superposition of two  different mass distributions $\ket{\DM_1}+\ket{\DM_2}$, as indicated by the blue and red configurations; the distributions can differ in mass $M_{1,2}$, size $1/\mu_{1,2}$, and center of mass location ${\bf L}_{1,2}$. As the probe particle passes by, the gravitational interaction causes it to also evolve into a quantum superposition of states $\ket{\psi_1}$ and $\ket{\psi_2}$, and thus it becomes entangled with the DM; see Eq.~(\ref{psiIF}).}
\label{fig:setup}
\end{figure}

When this DMSCS interacts with its environment, then this sub-system and the environment will co-evolve into an entangled state, one in which the state of the environment would depend on the states in the superposition, and thus the environment would also be in a superposition. We can denote the state of the environment as $\ket{\psi}$ and the state of the composite system as $\ket{\Psi}$. The initial ($\ket{\Psii}$) and final ($\ket{\Psif}$) states of the composite system are given by
\beq 
\ket{\Psii}=(\ket{\DM_1}+\ket{\DM_2})\ket{\psi},\,\,\,\,\,\,\,\,\,\,\ket{\Psif}=\ket{\DM_1}\ket{\psi_1}+\ket{\DM_2}\ket{\psi_2}
\label{psiIF}\eeq
where $\ket{\psi_1}$ is the state of the environment after interacting with $\ket{\DM_1}$, $\ket{\psi_2}$ is the same for $\ket{\DM_2}$. We ignore the backreaction of the probe particle on the DM; this will be an excellent approximation for the cases of interest, since the DMSCS will be much heavier than the probe particle, which will typically be an elementary particle, atom or molecule (see ahead to Eq.~(\ref{CharacteristicMmu})). Figure~\ref{fig:setup} gives a schematic of this entanglement process, where the role of the environment is played by a probe particle that passes through the DMSCS.

\subsection{Density Matrix}

Given a complete description of the state of the system, a more intuitive description of the probabilities associated with measurement outcomes is given by the density matrix $\hat{\rho}$. For a pure state it is defined as
\beq\hat{\rho}\equiv \ket{\Psi}\bra{\Psi}\eeq
For the combined description of the DM and its environment after interacting, the density matrix is given by
\beq \hat{\rho}=\Big(\ket{\DM_1}\ket{\psi_1}+\ket{\DM_2}\ket{\psi_2}\Big)\Big(\bra{\DM_1}\bra{\psi_1}+\bra{\DM_2}\bra{\psi_2}\Big)\eeq

Typically, the environment is made up of an extremely large number of particles and in practice this is far too many to keep track of in any detail. Thus, the statistical arrangement of the DM is effectively described by the so-called reduced density matrix, where the degrees of freedom of the environment have been traced out. This corresponds to ``ignoring'' those degrees of freedom and only ever keeping track of the states of the DM. This reduced density matrix is given by
\beq
\hat{\rho}_{\mbox{\tiny{red}}}=\mbox{Tr}_{\ket{\psi}}[\hat{\rho}]
=\ket{\DM_1}\bra{\DM_1}+\braket{\psi_2}{\psi_1}\ket{\DM_1}\bra{\DM_2} +\braket{\psi_1}{\psi_2}\ket{\DM_2}\bra{\DM_1} +\ket{\DM_2}\bra{\DM_2}
\eeq
The components that mix $\ket{\DM_1}$ and $\ket{\DM_2}$ will be referred to as the ``off-diagonal'' elements of the density matrix, since they would be off-diagonal when expressing the matrix in terms of the generalized basis vectors $\{\ket{\DM_1},\ket{\DM_2}\}$. These elements of the density matrix correspond to the interference phenomena which are unique to Schr\"{o}dinger-cat-like superpositions. The existence of these components can be taken as an indication that the system is behaving as a coherent quantum mechanical system. If the off-diagonal components were to vanish, then all that is left is a classical probability distribution for making a measurement of the system with no interference phenomena. 
So it is the off-diagonal components, which we denote
\beq
\Q\equiv\braket{\psi_1}{\psi_2}
\eeq
that are a measure of the residual quantumness in the reduced system.

\subsection{Perturbative Expansion}\label{sec:unitarity}

The DM in the galaxy experiences interactions with nearby SM particles through gravity. Thus, if a region of DM is in a macroscopic superposition of two different density profiles, the nearby SM particles have a chance to decohere this state. In order to compute the rate of decoherence, we will describe the interaction through the scattering of a probe particle off of the DM. Since the particles involved are assumed to be non-relativistic, the interaction will be described by Newtonian gravity. Once we determine the scattering solution of a probe particle near the DMSCS, we can compute the relevant overlaps of wave functions (the quantity $\Q$) and compute the rate of decoherence.

We will refer to the un-scattered incoming state of the probe particle as $\ket{u}$. As the particle nears the DM overdensity, it will start to be perturbed such that the state will evolve into 
\beq
\ket{\psi}=\ket{u}+\ket{s}
\eeq
where $\ket{s}$ is a piece from {\em scattering}. 
Since the resultant state evolves through the unitary time-evolution of the Schr\"{o}dinger equation, the normalization of the state must be preserved. Thus we need
\beq
\braket{\psi}{\psi}=\braket{u}{u} + \braket{u}{s}+\braket{s}{u}+\braket{s}{s} = 1
\eeq
Since the initial state should also be normalized $\braket{u}{u}=1$, this requires
\begin{equation}
\Re[\braket{u}{s}] = -\frac{1}{2}\braket{s}{s}
\label{NormC}\end{equation}

We will consider a potential which is weak enough to be treated perturbatively in the usual scattering formalism. Thus we can expand the scattered correction $\ket{s}$ in powers of a parameter, $\lambda$, which is used to keep track of the order at which the potential appears (in our case, we can think of expanding in powers of Newton's gravitational constant $G$), although we will ultimately set $\lambda=1$ as it is just an expansion tool. The scattering term is expanded as
\begin{equation}
\ket{s}=\lambda\ket{s^{(1)}}+\lambda^2\ket{s^{(2)}}+\ldots
\end{equation}
We can then use Eq.~(\ref{NormC}) self-consistently order by order in the expansion, which implies 
\beq
\Re\Big[\braket{u}{s^{(1)}}\Big]=0,\,\,\,\,\,\,\,\,\, \Re\Big[\braket{u}{s^{(2)}}\Big]=-\frac{1}{2}\braket{s^{(1)}}{s^{(1)}}
\eeq

We can apply this to each piece of the probe particle's induced superposition $\ket{\psi_1}$ and $\ket{\psi_2}$. We find that the overlap to second order is
\ba
\braket{\psi_1}{\psi_2}\amp=\amp 1 -i\, \lambda \bigg(\Im\Big[\braket{u}{s_1^{(1)}}\Big]- \Im\Big[\braket{u}{s_2^{(1)}}\Big]\bigg)+\lambda^2 \bigg(\braket{s_1^{(1)}}{s_2^{(1)}}\nonumber\\
\amp-\amp\frac{1}{2}\Big\{\braket{s^{(1)}_1}{s^{(1)}_1}+\braket{s^{(1)}_2}{s^{(1)}_2}\Big\}-i \Im\Big[\braket{u}{s_1^{(2)}}\Big]+i \Im\Big[\braket{u}{s_2^{(2)}}\Big]\bigg)+\ldots\label{psi1psi2 short}
\ea
Hence in order to work to $\mathcal{O}(\lambda^2)$, it naively appears as though we need to work to second order in the scattering, since $\ket{s_1^{(2)}}$ and $\ket{s_2^{(2)}}$ appear on the second line.

However, in order to describe decoherence, we are only interested in the {\em magnitude} of the overlap, not its overall phase. We can parameterize the deviation by $\Delta$, defined by 
\beq
|\braket{\psi_1}{\psi_2}|^2=1-2\Delta
\eeq
By taking the absolute square of Eq.~(\ref{psi1psi2 short}), we readily obtain
\ba
\Delta \amp=\amp 
{\lambda^2\over2} \bigg(\braket{s^{(1)}_1}{s^{(1)}_1}+\braket{s^{(1)}_2}{s^{(1)}_2} - 2\,\Re\Big[\braket{s_1^{(1)}}{s_2^{(1)}}\Big]\bigg)
\nonumber\\ \amp -\amp {\lambda^2\over2} \bigg( \Im\Big[\braket{u}{s_1^{(1)}}\Big]-\Im\Big[\braket{u}{s_2^{(1)}}\Big]\bigg)^2
+\ldots\label{psi1psi2 squared}
\ea
Therefore, to calculate the correction from unity up to order $\lambda^2$, it is only necessary to calculate the correction $\ket{s}$ up to order $\lambda$. The imaginary part of the second-order correction $\braket{u}{s_i^{(2)}}$ to $\langle\psi_1|\psi_2\rangle$ does not contribute at this order to $\Delta$, and the real part of the second-order correction has been re-written in terms of the first order correction. From this expression, we simply take a square root to give the absolute value of the wave function overlap. Thus, we will be able to compute the magnitude of the overlap to second order in perturbation theory while employing only the first-order Born approximated solution in the scattering formalism.

\subsection{Decoherence Rate}\label{DecRate}

In our set-up, the overlap between the two scattered wave functions will be close to one, i.e., $\Delta\ll 1$. We can then write
\beq
|\braket{\psi_1}{\psi_2}|=\sqrt{1-2\Delta}\approx 1-\Delta
\eeq
Since $\Delta$ will typically be small, it will typically take many scattering events for the off-diagonal elements of the density matrix to go towards zero. For $N$ such independent events, the overlap of $\ket{\psi_1}$ (the probe particle state after scattering off one part of the DM state $|\mbox{DM}_1\rangle$) and $\ket{\psi_2}$  (the probe particle state after scattering off one part of the DM state $|\mbox{DM}_2\rangle$) will decrease the off-diagonal elements of the density matrix $N$ times through the product
\beq
\Q = \prod_{n=1}^N|\braket{\psi_1}{\psi_2}|_n
\eeq
For a fixed quantum superposition of the target, the overlap $|\braket{\psi_1}{\psi_2}|_n$ will depend on the {\em impact parameter} $b$; where for small $b$, one expects a relatively big effect, while for large $b$, we expect a small effect. To be explicit, we can denote our correction to the overlap with a $b$-subscript as $\Delta_b$. As we shall see, $\Delta_b$ will typically be extremely small (see ahead to Figure \ref{fig:wf}), so we can estimate
\beq
\Q = \prod_{n=1}^N (1-\Delta_b)\sim \exp\left(-\sum_{n=1}^N\Delta_b\right)
\eeq
Note that here we are indicating that the important feature of the $n^{th}$ scattering is the impact parameter $b$. So while we could extend out notation to read $\Delta_{b,n}$, it will suffice to use the abbreviated notation $\Delta_b$. 
The rate at which the off-diagonal elements of the density matrix go to zero, and hence the system is decohering, is determined by the derivative of the argument of the exponent of $\Q$
\beq
\Gamd=-{d\over dt}\ln\Q\approx{d\over dt}\sum_{n=1}^N\Delta_b
\eeq

We will assume that the probe particles are moving isotropically and have uniform number density $n$ and so they have a uniform distribution of impact parameters. If the particles are moving with a common velocity $v$, 
then the rate is given by
\beq
\Gamd=n\,v\int \! d^2b\,\Delta_b
\label{DecRate}\eeq

The corresponding decoherence time $\tid$ can be defined as the time it takes the off-diagonal elements to be reduced by a factor of $e$. This is the solution to
\beq
\int_0^{\tid} dt'\,\Gamd(t') = 1
\label{tdecdef}\eeq
In general the instantaneous decoherence rate $\Gamd$ can itself be a function of time; for example, if the local number density of probe particles changes or if the state evolves. We will return to this possibility later, as it can be especially important for a DMSCS that passes by the earth or if its width spreads. On the other hand, if the decoherence rate is constant, then the decoherence time is $\tid=1/\Gamd$. The basic goal for the remainder of the paper is to compute these rates for DM.

\section{Scattering Theory}\label{Scattering}

To discuss the scattering of a probe particle off of an overdense region of DM, which sets up a potential energy function $V({\bf x})$, we employ the non-relativistic quantum mechanical scattering theory. The Lippman-Schwinger equation gives the solution for the scattered wave function. We first solve the Lippman-Schwinger equation for a plane wave, and then the plane wave solution is used to construct a Gaussian wave packet (useful literature sources include Refs.~\cite{Sakurai,MurayamaWebpage,Norsen,Ishikawa,Karlovets:2015nva}). We will first analyze the system under the simplifying assumption of a large wave packet, and then we will perform a more general analysis.

The Lippman-Schwinger solution is written as an unscattered plane wave plus an outgoing scattered wave. At radii $r=|{\bf x}|$ much larger than the characteristic scale of the potential, which we shall denote $1/\mu$, the (un-normalized) plane wave solution's form is well known to be
\beq
\psi_{\bf k}(\x)=e^{i\k\cdot\x}+f(\k',\k)\frac{e^{ikr}}{r}\label{Lip4}
\eeq
 Here $f(\k',\k)$ is the scattering amplitude, $\k$ is the wave vector for the incoming plane wave, and $\k'$ is a vector of magnitude $|{\bf k}'|=|{\bf k}|$ and points in the direction of interest ${\bf k}\parallel{\bf x}$. As described in the previous section, our interest is a scattering target that is in a superposition of two states. This means there are in fact two forms of $f$, associated with two forms of $V$, which we will later denote $f_i$, ($i=1,2$). But we will suppress the index $i$ for now and only use it later when we explicitly form the overlap of the two states $|\psi_1\rangle$,\,$|\psi_2\rangle$. 
 
As described in the previous section, we will only need the scattered solution to first order in perturbation theory. But for the sake of completeness, we know that the general solution is defined recursively by
 \beq 
 f(\k',\k)=-\frac{m}{2\pi\hbar^2}\int\! d^3 x' \,e^{-i\k'\cdot\x'}V(\x')\,\psi_{\bf k}(\x')\label{ampdef}
\eeq
where to first order we simply replace $\psi_{\bf k}(\x')\to e^{i\k\cdot\x'}$ in the integrand.

\subsection{Gaussian Wave Packet}

We would like to now build a wave-packet and properly describe its time evolution, from an initial state far away from the overdensity, evolving through the overdensity, and then to late times where it is once again far away. For convenience we will consider a Gaussian wave-packet. However, as we will see, our results will not be sensitive to this choice.

Suppose at some early initial time $t_e$ the probe particle is a Gaussian of spatial width $\Delta x=d$, centered at ${\bf x}=\hbar\,{\bf k}\,t/m$, with a central momentum $\hbar {\bf k}$ and width $\Delta k=1/2d$. We also assume the particle is heading towards the DM overdensity with impact parameter ${\bf b}$ that should be orthogonal to the central momentum ${\bf b}\cdot{\bf k}=0$. The magnitude of the momentum distribution is therefore
\beq
|\psi_{\bf k}({\bf q})|=\left(8\pi d^2\right)^{3/4} e^{-(\q-\k)^2 d^2}
\eeq
with phase that depends on the impact parameter. Taking the Fourier transform and evaluating at the early time $t_e$ (meaning well before scattering), we have
\beq\psi(\x,t_e)=\psi_u({\bf x},t_e)=\left(8\pi d^2\right)^{3/4} \! \int\! {d^3q\over(2\pi)^3}\, e^{i\q\cdot\x} e^{-iE_qt_e/\hbar} e^{-(\q-\k)^2 d^2} e^{-i{\bf q}\cdot{\bf b}}
\label{psiinitial}\eeq
where the energy of each plane wave component is $ E_q=\frac{\hbar^2 q^2}{2m}$. Then using Eq~\ref{Lip4}, the solution that incorporates scattering to first order in the potential at some later time $t$ is 
\beq
\psi({\bf x},t)=\psi_u({\bf x},t)+\psi_s({\bf x},t)
\eeq
where $\psi_u({\bf x},t)$ is simply the unscattered part, which is given by Eq.~(\ref{psiinitial}) with $t_e\to t$. While $\psi_s({\bf x},t)$ is the scattered part, given by
\beq
\psi_s(\x,t)=\left(8\pi d^2\right)^{3/4}\!\int\! {d^3 q\over(2\pi)^3}\,  f(\q',\q)\frac{e^{iqr}}{r} e^{-iE_qt/\hbar}e^{-(\q-\k)^2 d^2} e^{-i{\bf q}\cdot{\bf b}}
\label{packetint}
\eeq

In Section \ref{GenDer} we will analyze this in full generality. But for now, it is useful to analyze this with the following simplifying approximations: we assume that the size of the wave packet is much larger than any other scales in the problem and that the superposition states have a similar center of mass
\beq
(i)\,\,\, d\gg 1/k,\,\,\,\,\,\,\,\,\,(ii)\,\,\, d\gg 1/\mu,\,\,\,\,\,\,\,\,\,(iii)\,\,\,d\gg b ,\,\,\,\,\,\,\,\,\,(iv)\,\,\,1/\mu\gg L
\label{123}\eeq
The first condition (i) ensures that the probe particle's wave packet is very narrow in momentum space, so that the spread in momenta is much smaller than the central momenta $\Delta k=1/2d\ll k$. This is a very reasonable and realistic assumption, and is ultimately an assumption we will focus on for most of the paper (although we will be able to build a result that does not rely on this in the next section). The second condition (ii) means that the size of the probe particle's wave packet is larger than the size of the overdensity. This is an unrealistic assumption for probe particles in the atmosphere, whose size is microscopic, while it may be possible for probe particles in the galactic halo, whose wave packets can spread a lot. Nevertheless condition (ii) is useful for gaining an intuitive understanding of the behavior; in the next section we will relax this assumption, and derive a result that does not rely on (ii) at all. The third condition (iii) will ultimately be eliminated by including corrections from non-zero impact parameter $b$ when we put together the full rate. The fourth condition (iv) is that the distance between the center of mass of the two objects $L=|{\bf L}_1-{\bf L}_2|$ (which is only relevant when we form the overlap) is small. However, its effects will be fully incorporated in Section \ref{GenDer}. 

In this regime, the scattered term's integrand is strongly peaked at $\q=\k$ due to the $\sim e^{-({\bf q}-{\bf k})^2d^2}$ factor, and $f$ is slowly varying in this regime (under assumption (ii)). So we may evaluate the scattering amplitude $f$ at the value ${\bf q}\to{\bf k},\,{\bf q}'\to{\bf k}'$ and bring it outside of the integral. Then by operating at small impact parameter for the time being, we have 
\beq
\psi_s(\x,t)\approx \left(8\pi d^2\right)^{3/4} f(\k',\k)\int \!{d^3 q\over(2\pi)^3}\frac{e^{iqr}}{r}e^{-iE_qt/\hbar}e^{-(\q-\k)^2 d^2}\label{packetintapprox}
\eeq
The integrals over $\q$ can be computed explicitly, giving the total solution
\beq
\psi(\x,t)=\left(8\pi d^2\right)^{3/4}\bigg(\frac{\pi}{d^2+i\hbar t/2m}\bigg)^{3/2} e^{-k^2 d^2} \bigg[ e^{-\frac{\x^2-4 i \k\cdot\x d^2 - 4\k^2 d^4}{4(d^2+i \hbar t/2m)}}+f(\k',\k){\varphi_s(r,t)\over r}\bigg]\\
\eeq
where $\varphi_s(r,t)$ is a dimensionless function defined in Appendix \ref{app:scf}. Its asymptotic values at early and late times are 
\ba
\mbox{Early times:}\,\,\,\,\,\,\,\,\varphi_s(r,t)\amp\to\amp 0 \,\,\,\,\,\,\,\,\,\,\,\,\,\,\,\,\,\,\,\,\,\,\,\,\,\,\,\,\,\,\,\,\,\,\,\,\,\,\,\,\,\,\,\,\,\,\,\,\,\,\,\,\,\,\,\,\,
(\mbox{and}\,\,\,\,|\langle\psi_1|\psi_2\rangle|^2\to 1)\\
\mbox{Late times:}\,\,\,\,\,\,\,\,\,\,\varphi_s(r,t)\amp\to\amp {2d^2k+ir\over 2d^2k} e^{\frac{m \left(2 d^2 k+ir\right)^2}{4 d^2 m+2 i \hbar t}} \,\,\,\,\,(\mbox{and}\,\,\,\,|\langle\psi_1|\psi_2\rangle|^2\to 1-2\Delta)
\ea
At early times ($t$ large and negative) $\varphi_s\to0$ exponentially fast, ensuring that the wave function is just a free Gaussian wave packet.  On the other hand, at late times, once the wave packet has fully passed the origin ($t\gg d/v=md/(\hbar k)$), then Erf$(...)\to1$ and Erfc$(...)\to0$ (see Appendix \ref{app:scf}), and $\varphi_s$ organizes into 
the above form. By taking the absolute value of the wave function squared, it is simple to check that the unscattered part $\psi_u({\bf x},t)=\langle{\bf x}|u\rangle$ is peaked at ${\bf x}=\hbar\,{\bf k}\,t/m$, while the scattered part $\psi_s({\bf x},t)=\langle{\bf x}|s\rangle$ is peaked at $|{\bf x}|=r=\hbar\,k\,t/m$. 

\subsection{Gravitational Scattering}\label{GS}

We imagine now that the probe particle is scattering off of a DMSCS which is a small overdensity surrounded by a small underdensity in the dark matter in the galaxy. The total DM density can be written
\beq
\rho_i({\bf x}) = \rho_0({\bf x})+\delta\rho_i({\bf x})
\eeq
where $\rho_0({\bf x})$ is some (background) value that is common to each member of the superposition and will not be of importance to us as we are only interested in the final {\em differences} in wave functions. 
Of importance to us is $\delta\rho_i(r)$ ($i=1,2$), which is a perturbation that differs between the 2 parts of the wave function; this gives rise to the macroscopic superposition.  For the case of axion dark matter (see Section \ref{Numerics}), the mass density profile will be established by a huge number of coherent axions. The probe particle will therefore scatter off this large collection of particles which make up each part of $\rho_0$ and $\delta\rho_i$.
We will focus our attention on perturbations that do not carry any monopole, i.e.,
\beq
\int d^3x\,\delta\rho_i({\bf x}) = 0
\eeq
So at large distances, this  fluctuation is unnoticeable. But at closer approaches, a probe particle (such as a proton) feels a local Newtonian gravitational potential and has a non-zero probability to be scattered. One might think that this is a contrived setup. However, to the contrary, these are precisely the forms of superpositions that would be the most robust against decoherence. If the monopole itself was in a superposition, then it would be even more readily decohered by distant probes, thus leaving this type of residual perturbation left over in a superposition.

We parameterize the mass distribution of the DMSCS by two primary quantities: (i) a characteristic mass scale $M$ (while the distribution may integrate to zero mass, this will be the characteristic positive mass of the overdensity, which is cancelled by an equal and opposite underdensity) and (ii) characteristic width $1/\mu$ (and later we will refer to a distance between the centers of mass of the states). The specific form is written as (we will again suppress the index $i$ and reinstate later as needed)
\beq 
\delta\rho({\bf x})\equiv M\,\mu^3\, \zeta(\mu r)\label{deltarho}
\eeq
where $\zeta(\mu\, r)$ is a dimensionless function which captures the spatial dependence of the mass density in terms of the dimensionless variable $\mu\,r$. Furthermore, we will also assume this perturbed density profile is spherically symmetric. This means that not only is there no monopole, but it has no higher order multipole moments either. Again this is likely to be the most robust form of perturbation against decoherence.
 
This density profile is related to the Newtonian gravitational potential $\Phi_N$ by the Poisson equation
\beq 
\nabla^2\Phi_N(r)=4\pi G\, \delta\rho = 4\pi G\, M\, \mu^3\, \zeta(\mu r)\label{poisson}
\eeq
where $G$ is Newton's gravitational constant and the scattering potential energy is
\beq
V(r)= m\,\Phi_N(r)
\eeq 
with $m$ the mass of the probe particle.

We have checked that for the parameters of interest here, the perturbative analysis is valid; see ahead to Figure \ref{fig:wf}. In particular, we will be considering such weak gravitational sources that the classical deflection of a typical probe particle $\Delta\theta_{class}\sim GM\mu/v^2$ is extremely small. 

So to compute the scattering amplitude based on this potential, we can employ the Born approximation by computing $f(\k',\k)$ as a series in powers of an appropriate expansion parameter, which can be taken to be $G$. The term that is first order in $G$ is found by taking the wave function on the right hand side of Eq.~(\ref{ampdef}) to be the incoming plane wave solution.
\beq 
f^{(1)}(\k',\k)=-\frac{m^2}{2\pi\hbar^2}\int d^3x' e^{i(\k-\k')\cdot\x'}\Phi_N(\x')\label{f1}
\eeq
Eq.~(\ref{f1}) is directly related to the Fourier transform of the potential with respect to the transfer momentum
\beq
\transferbf\equiv\k-\k'
\eeq
Thus, we can define the first order scattering amplitude in terms of the Fourier transform of $\zeta(\mu r)$. The Fourier transform can be written in terms of the usual spatial and wavenumber coordinates. However we will find it especially useful to switch to dimensionless coordinates (labelled with a hat on top) defined by scaling out the characteristic scale in the distribution $\mu$ as follows
\beq
\y\equiv\mu\, \x,\,\,\,\,\,\,\,\,\,\,\,\,{\bf \dv}\equiv{\transferbf\over\mu}
\eeq
We then define a Fourier transform of some function $F$ with respect to these dimensionless variables as
\beq
\hat{F}({\bf{\dv}})\equiv\int d^3\hat{x} \,F(\y) \,e^{i\bf{\dv}\cdot\y}
\eeq
Taking the (dimensionless) Fourier transform of Eq.~(\ref{poisson}) gives
\beq 
-\mu^2\,\dv^2\,\hat{\Phi}_N(\dv)=4\pi G\, M\, \hat{\zeta}(\dv)
\label{PoissonFT}\eeq 
where we have used the spherical symmetry to indicate that the Fourier transforms are only a function of the magnitude of the wavevector ${\bf \dv}$. This can be written as
\beq
\dv = |{\bf \dv}|={2k\over\mu}\sin(\theta/2)
\eeq
where we have taken the wavevector of the scattered wave to have the same magnitude as the incoming wave, $|\k'|=|\k|\equiv k$, and defined $\theta$ to be the angle between $\k$ and $\k'$. 

Inserting Eq.~(\ref{PoissonFT}) into Eq.~(\ref{f1}), the scattering amplitude at first order can be written as
\beq
f^{(1)}(\k',\k) =f(k,\theta)
=\frac{2GMm^2}{\hbar^2\,\mu^2\,\dv^2}\,\hat{\zeta}(\dv)\label{fzeta}
\eeq
where in the first equality we have indicated that for a spherically symmetric potential, the scattering amplitude has azimuthal symmetry, and thus the first order amplitude is only a function of $\theta$ for a given $k$.

\subsection{Wave Function Overlap}\label{overlapints}

We consider now a DMSCS that is a superposition of two different density distributions. We define for each sub-state a mass scale $M_{1,2}$, a characteristic size $1/\mu_{1,2}$, and a center of mass location ${\bf L}_{1,2}$. Thus, a pair of wave functions may be defined, corresponding to a probe particle being scattered by each of the two distributions; we refer to this pair as $|\psi_1\rangle$ and $|\psi_2\rangle$. As described in Section \ref{Decoherence}, the particle is then entangled with the DM. Our goal now is to compute the overlap of the two wave functions 
\beq \braket{\psi_1}{\psi_2}\equiv\int d^3 x\, \psi_1^*(\x)\psi_2(\x)\eeq
which controls the rate of decoherence from tracing out this degree of freedom.

We are interested in the value at late times after the scattering has taken place. Physically, this is like ``measuring'' the state of the DM by the scattered particle. The overlap of the two wave functions scattered off of the potentials will tend to a constant in time. This is a result of the potentials being short ranged and the unitarity of time evolution, which ensures that $\langle\psi_1|\psi_2\rangle$ is time independent when the states evolve under the {\em same} (free) Hamiltonian at late times. 

The various pieces of the overlaps may be written up to integrals in the angle $\theta$ from the ${\bf k}$ axis. 
For the leading order overlap between scattered and un-scattered, we find
\beq
\braket{u}{s_i}= {i\,k\over\pi} \int d^2\Omega\, e^{-4k^2d^2\sin^2(\theta/2)} f_i(k,\theta) \label{us2gen2}
\eeq
with index $i=1,2$, while for the leading overlap between scattered and scattered, we find
\beq
\braket{s_i}{s_j}=\frac{\sigma_{ij}(k)}{2\,\pi\,d^2}\label{s1s2gen}
\eeq
with indices $i=1,2$ and $j=1,2$, 
where we have introduced a $2\times2$ matrix of generalized cross sections $\sigma_{ij}$ defined as
\beq
\sigma_{ij}(k)\equiv\int d^2\Omega\,f_i^*(k,\theta)f_j(k,\theta) 
\label{sigmamatrix}\eeq
Here the diagonal elements $\sigma_{11}$ and $\sigma_{22}$ are standard cross sections, while the off-diagonal elements are of a truly quantum character, associated with the overlap between the different states.

We can express $f_i$ in terms of the dimensionless function $\hat{\zeta}$ that we introduced above, as in Eq.~(\ref{fzeta}). This allows us to extract out all the scales in the problem as follows
\ba
\braket{u}{s_i}\amp=\amp i \frac{4G M_i m^2 }{\hbar^2 k}\,\om_i\\
\braket{s_i}{s_j}\amp=\amp\frac{4 G^2 M_i M_j m^4 }{\hbar^4 k^2 d^2 \mu_i\mu_j}\,\ch_{ij}
\ea
where the dimensionless quantities $\om_i$ and $\ch_{ij}$ are given by the following
\ba
\om_i \amp\equiv\amp\int^{2k\over\mu_i}_0 {d\dv\over\dv}\, e^{-d^2\mu_i^2 \dv^2} \,\hat\zeta\left(\dv\right)\label{Omdef}\\
\ch_{ij}\amp\equiv\amp\int^{2k\over\sqrt{\mu_i\mu_j}}_0 \frac{d\dv }{\dv^3}\,\hat\zeta\!\left(\!\sqrt{\mu_i\over\mu_j}\,\dv\!\right)
\hat\zeta\!\left(\!\sqrt{\mu_j\over\mu_i}\,\dv\!\right)\label{Chidef}
\ea
Physically, we will mainly be interested in the case in which the probe particle's wavelength $\lambda=2\pi/k$ is much smaller than the size of the DM overdensity $1/\mu_i$, meaning that the end points of the integrals can be extended to infinity with good accuracy.  
Then, for $k\gg\mu_i$ and assuming  the hierarchy of scales $\mu_1/\mu_2$ is not too large or small, we will see that $\chi_{ij}$ is typically $\mathcal{O}(1)$.

\subsection{Special Case Decoherence Rate}
Having established the typical size of the contribution from the scattered-scattered overlap $\langle s_i|s_j\rangle$, we would like to estimate the relative contribution from unscattered-scattered $\langle u|s_j\rangle$ to the decoherence rate. To do so let us recall that here we are working under the simplifying assumption of (ii) which says $d\,\mu_i\gg 1$. This means the exponential factor in the integral of $\om_i$ suppresses the integral considerably. To estimate the integral, recall that we are interested in mass densities that carry no monopole. This means the corresponding Fourier transform $\delta\hat\rho(\dv)\propto\hat\zeta(\dv)$ must vanish in the $\dv\to 0 $ limit. Furthermore, for localization, the Fourier transform should involve even (positive) powers of $\dv$ only. This ensures the leading power should be (at least) quadratic in $\dv$ as follows
\beq
\hat\zeta\left(\dv\right)=\hat\zeta_0\,\dv^2+\ldots,\,\,\,\,\,\mbox{small}\,\,\dv
\label{zetasq}\eeq
and typically we will have a prefactor $\hat{\zeta}_0=\mathcal{O}(1)$. For $k\gg\mu_i$ the value of $\om_i$ is readily estimated as
\beq
\om_i\approx {\hat{\zeta}_0\over2 d^2\mu_i^2}
\eeq
This allows us to estimate the contribution to the overlap correction $\Delta$ from unscattered-scattered relative to scattered-scattered. Let us illustrate this with the $i=j=1$ case, giving
\beq
{(\Im\braket{u}{s_1})^2\over \braket{s_1}{s_1}} \approx {\hat{\zeta}_0^2\over\chi_{11}}{1\over d^2\mu_1^2}
\eeq
Since $\ch_{11}$ and $\hat{\zeta}_0$ are both $\mathcal{O}(1)$ in this regime, this shows that the contribution from unscattered-scattered is a relative factor of $\sim 1/(d\mu)^2$ compared to the contribution from scattered-scattered. Since the above analysis is only valid in the regime (ii) $d\gg1/\mu$, this means we can ignore this contribution, and it is only the scattered-scattered piece that contributes. 

Then inserting Eq.~(\ref{s1s2gen}) into the first line of Eq.~(\ref{psi1psi2 squared}) (with $\lambda=1$ and noting that the second line is ignorable), we find that the correction to the overlap wave function has the following form in terms of cross-sections
\beq
\Delta_0={1\over2\pi d^2}\left[{1\over2}\sigma_{11}+{1\over2}\sigma_{22}-\Re\,\sigma_{12}\right]
\label{Delta0}\eeq
where we have written $\Delta\to\Delta_0$ with a subscript 0 to indicate that this was done for zero impact parameter $b=0$. For the case at hand of gravitational scattering, the form that makes the scales in the problem manifest (recalling that $\chi_{ij}$ is typically $\mathcal{O}(1)$) is the following
\beq
\Delta_0 = \frac{2G^2 m^4 }{\hbar^4 k^2 d^2}\left[{M_1^2\over\mu_1^2}\ch_{11}+{M_2^2\over\mu_2^2}\ch_{22}-2{M_1M_2\over\mu_1\mu_2}\ch_{12}\right]
\label{Delta02}\eeq

Now we would like to include the contribution from non-zero impact impacter $b$. Since the Gaussian wave packet is very large in the present analysis $d\gg 1/\mu$ and $d\gg L$ (see next subsection for a more general treatment), we can say that the probability that the particle passes through the overdensity is given by the square of the wave function at closest approach $r=b$. This means
\beq
\Delta_b=\Delta_0\,{|\psi(b)|^2\over|\psi(0)|^2}=\Delta_0\,e^{-b^2/(2 d^2)}
\eeq
The integral over impact parameter is then very simple, giving
\beq
\int d^2b\,\Delta_b =2\pi\,d^2\,\Delta_0
\eeq
Note that this cancels the factor of $1/(2\pi d^2)$ in Eq.~(\ref{Delta0}). This means the final decoherence rate from Eq.~(\ref{DecRate}) takes on the very intuitive form
\beq
\Gamd = {1\over2}n\,v\left[\sigma_{11}+\sigma_{22}-2\,\Re\,\sigma_{12}\right]
\label{Gammasigma}\eeq
This is very closely related to the rate of scattering of particles off a single target $\Gamma_{sc}=n\,v\,\sigma$ and is similar to decoherence of a single point particle whose center of mass is in a superposition; see  Refs.~\cite{Joos:1984uk,Gallis1990,Diosi1995,Giulini:1996nw,Kiefer:1997hv,Dodd:2003zk,Hornberger,Schlosshauer:2003zy,SchlosshauerBook,HornbergerIntrp,Schlosshauer:2019ewh}.  
However the difference is that we are considering the rate of decoherence of a superposition of two different states for the DM target. We see that it organizes into this nice generalization (\ref{Gammasigma}), involving the generalized cross section $\sigma_{ij}$. 
For the case of gravitational scattering, we then have
\beq
\Gamd = \frac{4\pi G^2 m^4 n\, v}{\hbar^4 k^2}\left[{M_1^2\over\mu_1^2}\ch_{11}+{M_2^2\over\mu_2^2}\ch_{22}-2{M_1M_2\over\mu_1\mu_2}\ch_{12}\right]
\label{Gammagravity}\eeq
These are some of our primary results, but they will be generalized in the next subsection.

\subsection{General Case Decoherence Rate}\label{GenDer}

Note that the size of the wave packet $d$ has dropped out of the final results above. This makes it tempting to think that the result is true even if we relax the assumption (ii) $d\gg 1/\mu$, which is in fact true, as we shall show. However, for greater generality, we will now also relax the assumption (iv) $1/\mu\gg L$, allowing a finite separation between the centers of mass, and show that a new feature appears for the off-diagonal terms $\sigma_{12}$. 

In the above analysis with $d\gg 1/\mu$ we saw that the unscattered-scattered contribution was ignorable. We have also checked explicitly that it is ignorable in the opposite regime of $d\ll 1/\mu$. In fact, one can check, using analysis similar to what we present below, that it is irrelevant in any regime and therefore we can purely focus on the scattered-scattered contribution. 

We will now proceed with a much more general analysis, in which we don't assume (ii), (iii), or (iv) of Eq.~(\ref{123}) (though we will assume $d\gtrsim 1/k$ so that there is at least some rough localization of momenta). This means we return to the general expression for the scattered wave function in Eq.~(\ref{packetint}). As the primary building block of the decoherence rate, let us now define
\beq
S_{ij} \equiv \int d^2b\,\langle s_i|s_j\rangle = \int d^2b\int d^3 x\,\psi_{si}^*({\bf x},t)\psi_{sj}({\bf x},t)
\eeq
which is useful to construct $\int d^2b\,\Delta_b={1\over2}(S_{11}+S_{22}-2\Re S_{12})$ and then $\Gamd$ using Eq.~(\ref{DecRate}). 

For complete generality we can assume that the wave packet is not merely a Gaussian, but has some momentum distribution of $|\psi_{\bf k}({\bf q})|e^{-i{\bf b}\cdot{\bf q}}$, where again the ${\bf k}$ subscript just indicates it may have characteristic momentum ${\bf k}$. We will also allow the centers of mass of the two parts of the superposition to be different; we will label their positions ${\bf L}_i$.

The first order scattering theory, tells us that the integral at large distances may be written as
\ba
S_{ij} = \int d^2b\!\int d^3x\!\int \!{d^3q\over(2\pi)^3}\!\int \!{d^3\tilde{q}\over(2\pi)^3}
\!\!\!\!\amp\amp {e^{-i(qr_i-\tilde{q}r_j)}\over r^2} e^{i(E_q-E_{\tilde{q}})t/\hbar}
\, e^{i(({\bf b}-{\bf L}_i)\cdot{\bf q}-({\bf b}-{\bf L}_j)\tilde{\bf q})}
\nonumber\\
\amp\amp\times\, f_i^*({\bf q}',{\bf q})f_j(\tilde{\bf q}',\tilde{\bf q})\, |\psi_{\bf k}({\bf q})\psi_{\bf k}(\tilde{\bf q})| 
\ea
where we have defined the distance from the $i^{th}$ center of mass ${\bf L}_i$ to a distant radial point as
\beq
r_i \equiv |{\bf x}-{\bf L}_i|\approx r-\hat{\bf x}\cdot{\bf L}_i
\eeq
We can now immediately do the integral over impact parameter
\beq
\int d^2b\, e^{-i{\bf b}\cdot({\bf q}-\tilde{\bf q})}=(2\pi)^2\delta^2({\bf q}_\perp-\tilde{\bf q}_\perp)
\eeq
where ${\bf q}_\perp$ means the component of the momentum vector that is orthogonal to the central axis, i.e., ${\bf q}_\perp\cdot{\bf k}=0$ since the impact parameter vector satisfies ${\bf b}\cdot{\bf k}=0$. The integral over space can be partially performed as follows
\beq
\int d^3 x{e^{-i(q-\tilde{q})r}\over r^2} = \int d^2\Omega\int_0^\infty dr\,e^{-i(q-\tilde{q})r}\to\int d^2\Omega\,(2\pi)\delta(q-\tilde{q})
\eeq
where we have used the fact that at late times, we know that the full integral is dominated by the wave packets at large distances from the origin, allowing us to extend the integral over $r$ from $-\infty$ to $\infty$, which gives a delta-function. 

Altogether the above pair of integrals over impact parameter ${\bf b}$ and radius $r$ give the 3-dimensional delta function
\beq
(2\pi)^3\delta^2({\bf q}_\perp-\tilde{\bf q}_\perp)\delta(q-\tilde{q})\approx (2\pi)^3\delta^3({\bf q}-\tilde{\bf q})
\eeq
where we have used the fact that if ${\bf q}_\perp=\tilde{\bf q}_\perp$ and $q=\tilde{q}$, then we must have $q_\parallel=\tilde{q}_\parallel$ {\em or} $q_\parallel=-\tilde{q}_\parallel$. But so long as we have any reasonable localization of the wave packet, we will need ${\bf q}\approx\tilde{\bf q}\approx{\bf k}$ and therefore we can ignore the option where the parallel momenta are opposite; this option will be exponentially suppressed.

We can then immediately do the integral over $\tilde{\bf q}$, giving
\beq
S_{ij} = \int\! d^2\Omega \!\int \!{d^3q\over(2\pi)^3}\,f_i^*({\bf q}',{\bf q})f_j({\bf q}',{\bf q})\,|\psi_{\bf k}({\bf q})|^2
e^{-i({\bf q}-{\bf q}')\cdot{\bf L}_{ij}}
\label{SijL}\eeq
where ${\bf L}_{ij}\equiv{\bf L}_i-{\bf L}_j$ is the separation vector between the $i^{th}$ and $j^{th}$ parts of the superposition (so ${\bf L}_{11}={\bf L}_{22}=0$ and only ${\bf L}_{12}$ can be non-zero). The presence of the vector  ${\bf L}_{ij}$ between the centers of mass breaks the axisymmetry of the problem, leaving this moderately complicated expression. We can proceed by noting that we expect the probe particles to be moving isotropically with central wavevectors ${\bf k}$ having uniformly distributed orientations (this may not be exactly applicable for DM near the earth, where the atmosphere of the earth picks a preferred direction relative to the cosmic wind, but it will suffice for our purposes). Hence we average over the direction of ${\bf k}$ and define
\beq
\langle S_{ij}\rangle_{\hat{\bf k}} \equiv {1\over4\pi}\int\! d^2\Omega_{\bf k} \, S_{ij}({\bf k})
\eeq
This averaging only affects the $|\psi_{\bf k}({\bf q})|^2$ factor in Eq.~(\ref{SijL}) giving the function $P_k(q)\equiv \langle |\psi_{\bf k}({\bf q})|^2\rangle_{\hat{\bf k}}$ which only depends on the magnitude of the wavevectors $k=|{\bf k}|,\,q=|{\bf q}|$. 

We can then perform the integral over $d^2\Omega_q$ (the angular part of $d^3 q$) to obtain an even more general form for the generalized matrix of cross sections 
\beq
\tilde{\sigma}_{ij}(q)\equiv\int \!d^2\Omega\,f_i^*(q,\theta)f_j(q,\theta)\,j_0(2q L_{ij}\sin(\theta/2))
\eeq
where $j_0(z)\equiv \sin(z)/z$ is the sinc-function and $L_{ij}\equiv|{\bf L}_{ij}|$. Note that $\tilde{\sigma}_{11}=\sigma_{11}$ and $\tilde{\sigma}_{22}=\sigma_{22}$, since $j_0(0)=1$, matching what we defined earlier in Eq.~(\ref{sigmamatrix}) (here evaluated at $q$ rather than $k$). While the off-diagonal term $\tilde{\sigma}_{12}$ differs from $\sigma_{12}$ if we have a non-zero separation in the centers of mass $L=L_{12}$. 

Altogether this gives the final result
\beq
\langle S_{ij}\rangle_{\hat{\bf k}} = {1\over 2\pi^2}\int_0^\infty dq\,q^2\,\tilde\sigma_{ij}(q)\,P_k(q)
\eeq
Hence the (angle-averaged) $S_{ij}$ is the average value of the generalized cross section, weighted by the wave packet's momentum distribution. If we now assume that the distribution is well localized in momentum space (i) $\Delta k\ll k$, then the total cross section can be taken outside of the integral. We then obtain
\beq
\langle S_{ij}\rangle_{\hat{\bf k}} \approx \tilde\sigma_{ij}(k)
\eeq

Having established this result, we can insert this into Eqs.~(\ref{psi1psi2 squared},\,\ref{DecRate}) and we obtain a generalization of the earlier results for the decoherence rate in Eqs.~(\ref{Gammasigma},\,\ref{Gammagravity}) with the replacements $\sigma_{ij}\to\tilde\sigma_{ij}$ and $\chi_{ij}\to\tilde\chi_{ij}$ defined accordingly as
\beq
\tilde\ch_{ij}\equiv\int^{2k\over\sqrt{\mu_i\mu_j}}_0 \frac{d\dv }{\dv^3}\,\hat\zeta\!\left(\!\sqrt{\mu_i\over\mu_j}\,\dv\!\right)
\hat\zeta\!\left(\!\sqrt{\mu_j\over\mu_i}\,\dv\!\right) j_0(L_{ij}\sqrt{\mu_i\mu_j}\,\dv)\label{Chidef2}
\eeq
(note $\tilde\chi_{11}=\ch_{11}$ and $\tilde\ch_{22}=\ch_{22}$). 
Now our results are in much greater generality, as they are independent of whether the wave packet size $d$ is larger or smaller than the size of the potential $1/\mu$, they incorporate the integration over impact parameter rigorously, and they allow for a finite separation between the centers of mass $L$.

\section{Application to Dark Matter}\label{Numerics}

In this section, we take the DM to be light bosons, whose best motivated examples are axions \cite{Peccei:1977hh,Weinberg:1977ma,Wilczek:1977pj,Preskill:1982cy,Abbott:1982af,Dine:1982ah,AxionBook} and axion-like particles \cite{Jaeckel:2010ni}. In this case the DM particles can be extremely light, which gives rise to rather large de Broglie wavelengths, high occupancy states. If they evolve into quantum superpositions, then they would be macroscopic Schr\"{o}dinger-cat-like states. We then use ordinary matter probe particles in either the galactic halo or the earth's atmosphere to act as its environment.

\subsection{Mass and Length Scale Parameterization}\label{sec:params}

The size of the statistical overdensities in the axion field is set by the typical de Broglie wavelength of the axion
\begin{equation}\lambda_{dB}\equiv\frac{2\pi \hbar}{m_a v_a}
\end{equation}
where $m_a$ and $v_a$ are the axion mass and the typical speed of the axion in the galaxy, respectively. The characteristic width of the overdensity $1/\mu_i$ for statistical fluctuations is plausibly of order the de Broglie wavelength. We will parameterize it 
by the dimensionless parameter $\beta_i$, defined through
\beq
1/\mu_i=\beta_i\,\lambda_{dB}
\eeq
Similarly, the characteristic density of a statistical overdensity is of the order of the typical dark matter density itself $\rhoDM$. We can parameterize the corresponding mass by a dimensionless parameter $\alpha_i$, defined through
\begin{equation}
M_i=\alpha_i\frac{4\pi}{3}\rhoDM\lambda_{dB}^3
\end{equation}
Finally, the distance between the superpositions $L$ depends on more detailed dynamics. We can parameterize the distance between the centers of mass $L$ in terms of the widths of the DMSCS profiles and another dimensionless parameter $\gamma$ as
\beq
L = \gamma/\sqrt{\mu_1\mu_2}
\eeq
This leads to multiple forms for the superposition, which we organize as options (I), (II), (III) (though a general combination of possibilities is allowed).

(I) One possible superposition is that we fix the density to be the galactic value, meaning that both the size and mass are dictated by the same parameter $\beta_i$. This means we set 
$\alpha_i=\beta_i^3$. 
Here we also choose to set the two states to have the same center of mass $L=0$; so the two states are characterized entirely by $\beta_1\neq \beta_2$.

(II) Alternatively, we may choose to examine the case where the length scales for both states are the same ($\mu\equiv\mu_1=\mu_2$), in which case there is only one $\beta$, and again the same center of mass $L=0$. Then the states are purely distinguished by their mass parameters $\alpha_1\neq\alpha_2$. 

(III) Finally, we can consider the case where both the mass and sizes are the same ($M_1=M_2$, $\mu\equiv\mu_1=\mu_2$), and the superpositions are simply distinguished by a non-zero separation between their centers of mass $L$. The superposition is then specified by the value of $\gamma=L\,\mu$.

Also, for the sake of clarity, to describe probe particle properties, we will replace $n\to n_p$, $v\to v_p$, $m\to m_p$, $k\to m_p v_p/\hbar$ in our earlier formulae. We can also write the number density of probe particles in terms of mass density as $n_p=\rho_p/m_p$. 

A form for the mass distribution must be specified to calculate $\Delta$ precisely. For the DMSCS, we can choose a variety of mass distributions so long as they result in Newtonian potentials that have a finite range of support (scale $1/\mu_i$). In terms of the above parameters and the integrals $\ch_{ij}$, the decoherence rate is given by
\beq
\Gamd = {2^{14}\pi^{11}G^2\hbar^6m_p\rhoDM^2\rho_p \over 9 m_a^8 v_a^8 v_p}\beta_1\beta_2\,\alpha_1\alpha_2\,X
\label{Gam1}\eeq
where we have introduced a dimensionless function $\potl$ that is in general a function of the relative mass, size, shape, and separation of the two parts of the superposition, it is given by
\beq 
\potl = \frac{\beta _1\alpha_1}{\beta_2\alpha_2}  \ch _{11}+\frac{\beta _2\alpha_2}{\beta_1\alpha_1} \ch _{22}-2 \tilde\ch _{12}(\gamma)
\label{Xdefmg}  \eeq
Unless we take extreme cases, then $\potl$ is often an $\mathcal{O}(1)$ number.

In case (I), $\potl$ depends on only the ratio $\beta_1/\beta_2$ and choice of distribution as follows
\beq 
\mbox{(I)}\,\,\,\,\,\,\,\,\potl = \frac{\beta _1^4}{\beta_2^4}  \ch _{11}+\frac{\beta _2^4}{\beta_1^4} \ch _{22}-2 \ch _{12}
\label{Xdef}  \eeq
In case (II), where $\mu=\mu_1=\mu_2$, and the mass is distinguished by the parameters $\alpha_1,\,\alpha_2$, the function $\potl$ is given by
\beq
\mbox{(II)}\,\,\,\,\,\,\,\,\potl = \ch\left(\sqrt{\alpha_1\over\alpha_2}-\sqrt{\alpha_2\over\alpha_1}\right)^2
\label{Gam2}\eeq
Finally, in case (III), where $\mu=\mu_1=\mu_2$ and $M_1=M_2$ and the mass is distinguished by its separation in center of mass $L$, the function $\potl$ is given by
\beq
\mbox{(III)}\,\,\,\,\,\,\,\,\potl = 2(\ch_{11}-\tilde\ch_{12}(\gamma))
\label{Gam3}\eeq
where $\tilde\ch_{12}(\gamma)$ depends on the ratio of separation $L$ to size $1/\mu$.

\begin{figure}[t]
\centering
\includegraphics[width=.65\linewidth]{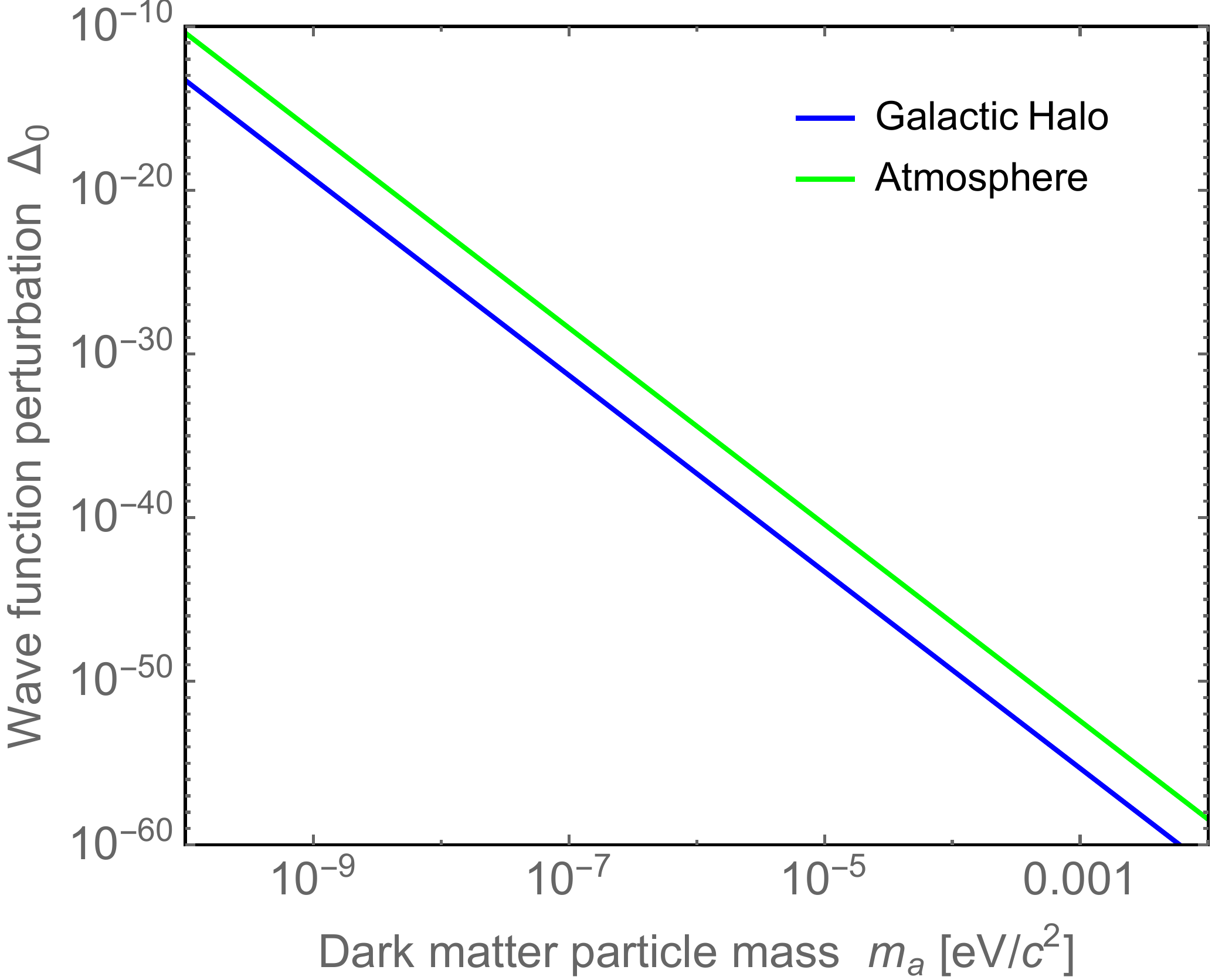}
\caption{The wave function perturbation $\Delta_0$ (at zero impact parameter) as a function of DM particle mass $m_a$
(with $\beta_1=1/(\lambda_{dB}\,\mu_1)=1$ and $M_2=0$, so $\Delta=\langle s_1|s_1\rangle/2$). The blue curve is for probe particles of protons (hydrogen), as relevant to the halo, while the green curve is for probe particles of $N_2$ molecules, as relevant to the earth's atmosphere. Further details are explained ahead in Section \ref{QR}. The fact that $\Delta_0\ll 1$ means that we can self-consistently use the perturbation theory that we have employed in this work.}
\label{fig:wf}
\end{figure}

We note that for sufficiently tiny axion masses, we will eventually obtain a large value of $\Delta$. In such a regime the perturbative analysis is breaking down. In Figure~\ref{fig:wf} we plot the case of zero impact parameter $\Delta_0$, as it is largest. For small probe particle wave packets, this is given by Eq.~(\ref{Delta02}) with the interchange $1/d^2\to(2\mu)^2$ for the upcoming Gaussian density profile (discussed in Section~\ref{sec:examplemass}). As the plot shows, for the parameters of interest discussed in the following sub-sections, this breakdown does not occur, i.e., $\Delta$ satisfies $\Delta\ll 1$ (in the figure we have set $M_2=0$, so $\Delta=\langle s_1|s_1\rangle/2$ and is therefore a direct test of the validity of perturbation theory). This self-consistently means that each probe particle is traveling in a nearly straight line, almost completely unperturbed, except for an extremely tiny deflection, reflection, and time delay. Since there is inevitable spreading of the wave function, the overlap is constant at late times. 

\subsection{Example Density Distributions}\label{sec:examplemass}

We now use these results to determine the decoherence rate of the DMSCS in the halo of the galaxy and at the earth. 
The shape of the overdensity must be specified for a precise quantitative result; however we shall see that our results are relatively insensitive to this choice.

As discussed in Section \ref{GS} there are some basic conditions that the mass distributions are expected to satisfy so as to be the most robust against decoherence and therefore of the most interest. We can summarize these properties as (i) being localized, (ii) having a vanishing monopole, and (iii) being spherically symmetric. Note that together these imply all multipole moments vanish. 

Two illustrative examples of density distributions that exhibit these properties are a ``Gaussian" profile ($\sim e^{-\mu^2 r^2}$) and a ``Yukawa" profile ($\sim e^{-\mu r}$), as follows
\ba 
\delta\rho_G(r)\amp=\amp M \mu^3 \!\left(1-\frac{2\mu^2}{3}r^2\right) e^{-\mu^2 r^2}\,\,\,\,\,\,(\mbox{Gaussian})\label{deltarhog}\\
\delta\rho_Y(r)\amp=\amp M e^{-\mu r}\!\left(\delta^3({\bf x})-\frac{\mu^2}{4\pi r}\right)\,\,\,\,\,\,\,\,\,\,\,\,(\mbox{Yukawa})
\ea
(and the centers can be shifted to ${\bf L}_i$ as appropriate). The Gaussian case is rather more realistic, as the distribution is smooth. The Yukawa case is rather unrealistic as it has a delta-function piece, which is not typical of light axion DM which has a large de Broglie wavelength, and also becomes very negative. Nevertheless, they will both be useful to illustrate the basic ideas, and despite their apparent large differences, the final results will turn out to be similar.

The details of the calculations involving each of these mass distributions, which amount to computing the integrals $\ch_{ij}$, are given in Appendices~\ref{app:gauss} and~\ref{app:yuk}. In this section, we summarize the results. Their dimensionless Fourier transforms are readily found to be
\ba
\hat\zeta_G(\dv)\amp=\amp{\pi^{3/2}\over6}\,\dv^2\,e^{-\dv^2/4}\,\,\,\,\,\,\,(\mbox{Gaussian})\\
\hat\zeta_Y(\dv)\amp=\amp{\dv^2\over 1+\dv^2}\,\,\,\,\,\,\,\,\,\,\,\,\,\,\,\,\,\,\,\,\,\,\,\,\,\,(\mbox{Yukawa})
\ea
Note that in both cases the Fourier transforms vanish as $\dv\to0$, since there is no monopole, and they are both analytic in powers of $\dv^2$, since the source is localized, as we mentioned earlier in Eq.~(\ref{zetasq}) (we see $\hat{\zeta}_{0G}=\pi^{3/2}/6$ and $\hat{\zeta}_{0Y}=1$, which are $\mathcal{O}(1)$ as anticipated).

For these distributions, we can compute the function $\potl$ explicitly. 
In case (I), we use the definitions in Eqs.~(\ref{Chidef},\ref{Xdef}) and find the explicit form of this function for $k\gg\mu$ is
\ba
\mbox{(I)}\,\,\,\,\,\,\,\,\potl_G\amp=\amp\frac{\pi^3}{36}   \left(\frac{\beta _1^4}{\beta _2^4} +\frac{\beta _2^4}{\beta _1^4} -\frac{4 \beta _2\beta_1}{{\beta
   _2^2}+{\beta _1^2} }\right)\,\,\,\,\,\,\,\,\,\,\,(\mbox{Gaussian})\label{chigauss}\\
\mbox{(I)}\,\,\,\,\,\,\,\,\potl_Y\amp=\amp\frac{1}{2}\frac{\beta _1^4}{\beta _2^4}+\frac{1}{2}\frac{\beta _2^4}{\beta _1^4}-\frac{2 \beta _2 \beta_1\ln \left(\frac{\beta _1}{\beta _2}\right)}{{\beta _1^2}-{\beta _2^2}}\,\,\,\,\,\,(\mbox{Yukawa})\label{chiyuk}
   \ea
One can check that, 
up to an overall numerical factor of $\pi^3/18\approx 1.7$, the value of the function $\potl$ is almost the same for each potential, and thus the decoherence rate is not very sensitive to the choice of the density distribution. Note that for $\beta_2\approx\beta_1$ we can approximate the $\potl$ function as
\beq
\mbox{(I)}\,\,\,\,\,\,\,\,\potl\approx \cc {(\beta_1-\beta_2)^2\over\beta^2}\,,\,\,\,\,\,\,\,\,\,\,(\mbox{for}\,\,\,\beta_2\approx\beta_1)
\label{Xapprox}\eeq
where $\cc$ is a pre-factor depending on the distribution: $\cc_G=17\pi^3/36$, $\cc_Y=49/6$. So, naturally, the corresponding decoherence rate will descrease (quadratically) as we take the two DM states closer and closer to each other. 

Also, recall that in case (III), the dimensionless separation $\gamma=L\,\mu$ plays a central role. By using Eqs.~(\ref{Chidef2},\ref{Gam3}) the function $\potl$ is given in terms of $\gamma$ as follows
\ba
\mbox{(III)}\,\,\,\,\,\,\,\,\potl_G\amp=\amp{\pi^3\over18}\left(1-{\sqrt{2}\over\gamma}\,D_+\!\left(\gamma\over\sqrt{2}\right)\right) 
\,\,\,\,\,\,\,\,\,\,\,\,(\mbox{Gaussian}) \\
\mbox{(III)}\,\,\,\,\,\,\,\,\potl_Y\amp=\amp 1-{\sqrt{\pi}\over2}\, G_{0,1}^{2,1}\!\left({\gamma^2\over4}\Bigg{|}
\begin{array}{cc} 0\\ 0\,\, 1\, {-1\over2}\end{array}\! \right) \,\,\,\,\,(\mbox{Yukawa})
\ea
where we have written the answers in terms of $D_+$, the so-called Dawson's integral function, and $G$, the so-called MeijerG function. For small separations, $\gamma=\mu\,L\ll 1$, we can expand these to obtain
\beq
\mbox{(III)}\,\,\,\,\,\,\,\,\potl\approx \ccc\,\gamma^2\,,\,\,\,\,\,\,\,\,\,\,(\mbox{for}\,\,\,\gamma\ll 1)
\label{XapproxIII}\eeq
where $\ccc$ is a pre-factor. In the Gaussian case it is given by $\ccc_G=\pi^3/54$. While in the Yukawa case it is in fact not analytic; it is given by $\ccc_Y=(3\ln(1/\gamma)+4-3g)/9$ (where $g\approx0.577$ is the Euler-Mascheroni constant). This logarithm is presumably an artifact of the spike at the center of the Yukawa distribution and is not expected to be present for smooth distributions. 

\subsection{Quantitative Results}\label{QR}

We now quantify the decoherence rate of Eq.~(\ref{Gam1}) applied to the Gaussian potential using Eq.~(\ref{chigauss}). The results differ for the Yukawa potential by an overall factor $\pi^3/18\approx1.7$, and thus the results are very similar. The local DM density in the milky way galaxy is taken to be
\begin{equation}
\rhoDM \approx 0.4 \frac{\mbox{GeV}/c^2}{\mbox{cm}^3}
\end{equation}
(for a review, see Ref.~\cite{Read:2014qva}). A characteristic speed of particles is the galactic rotation speed at the solar circle of $v_0\approx 220\,\mbox{km}/$\,s \cite{Kuhlen:2012ft} and so we take $v_a\approx v_0$.

In displaying our upcoming results, we will often make reference to a representative mass for the axion of $m_a=10^{-6}$\,eV$/c^2$. In the particular case of the QCD axion, the relic abundance from the misalignment mechanism in the pre-inflationary scenario is known to be
\beq
\Omega_a\,h^2\approx \left(10^{-6}\,\mbox{eV}\over m_a\,c^2\right)^{\!1.165}\Theta_i^2
\eeq
(for a brief summary, see \cite{PDG2018}), 
where $\Theta_i$ is the initial misalignment angle, which can be $\mathcal{O}(1)$ or smaller. Hence one often considers axion masses of $m_a\sim \mbox{few}\times10^{-6}$\,eV$/c^2$ as plausible DM candidates, though lighter values of $m_a$ are allowed if $\Theta_i$ is small. In the post-inflationary scenario, moderately higher values of $m_a$ are preferred due to the production of axion-strings (and $\Theta_i^2$ gets averaged to an $\mathcal{O}(1)$ number). In general, though, the possible value of generic axion masses, including string theory axions and axion-like particles, can span orders of magnitude. 
In any case, we write the characteristic mass  and length scales as
\beq
M_i\sim 10^{16}\,\mbox{GeV}/c^2\!\left(10^{-6}\,\mbox{eV}\over m_a\,c^2\right)^{\!3} \alpha_i,
\,\,\,\,\,\,\,\,\mu_i^{-1}\sim 2\,\mbox{km}\!\left(10^{-6}\,\mbox{eV}\over m_a\,c^2\right)\beta_i
\label{CharacteristicMmu}\eeq

Now, as for the probe particles, we take them to be the diffuse free protons (hydrogen) which co-exist with the DM throughout the galaxy. The mass of the proton is $m_p=0.938$\,GeV$/c^2$, and we will take the speed of diffuse protons to also be set by the typical galactic speed $v_p\approx v_0$. More precisely, we could integrate our result over a Maxwell velocity distribution for $v_p$, but since Eq.~(\ref{Gam1}) only scales as a single power of $1/v_p$, this only leads to a small correction. Since we know ordinary matter only comprises a fifth as much as the DM total density, we set the proton (hydrogen) density in the halo to be 
\beq
\rho_p\approx 0.2\,\rhoDM
\eeq
(although the exact value would vary depending on how far out from the galactic disk one is considering). 
One could also use galactic photons as probe particles, although this would require a relativistic analysis. Naively, in this case the result in Eq.(\ref{Gam1}) should be just altered by $m_p\rho_p/v_p\to E_\gamma\rho_\gamma\,\,(c=1)$, which is much smaller.

 \begin{figure}[t]
\centering
\includegraphics[width=0.65\linewidth]{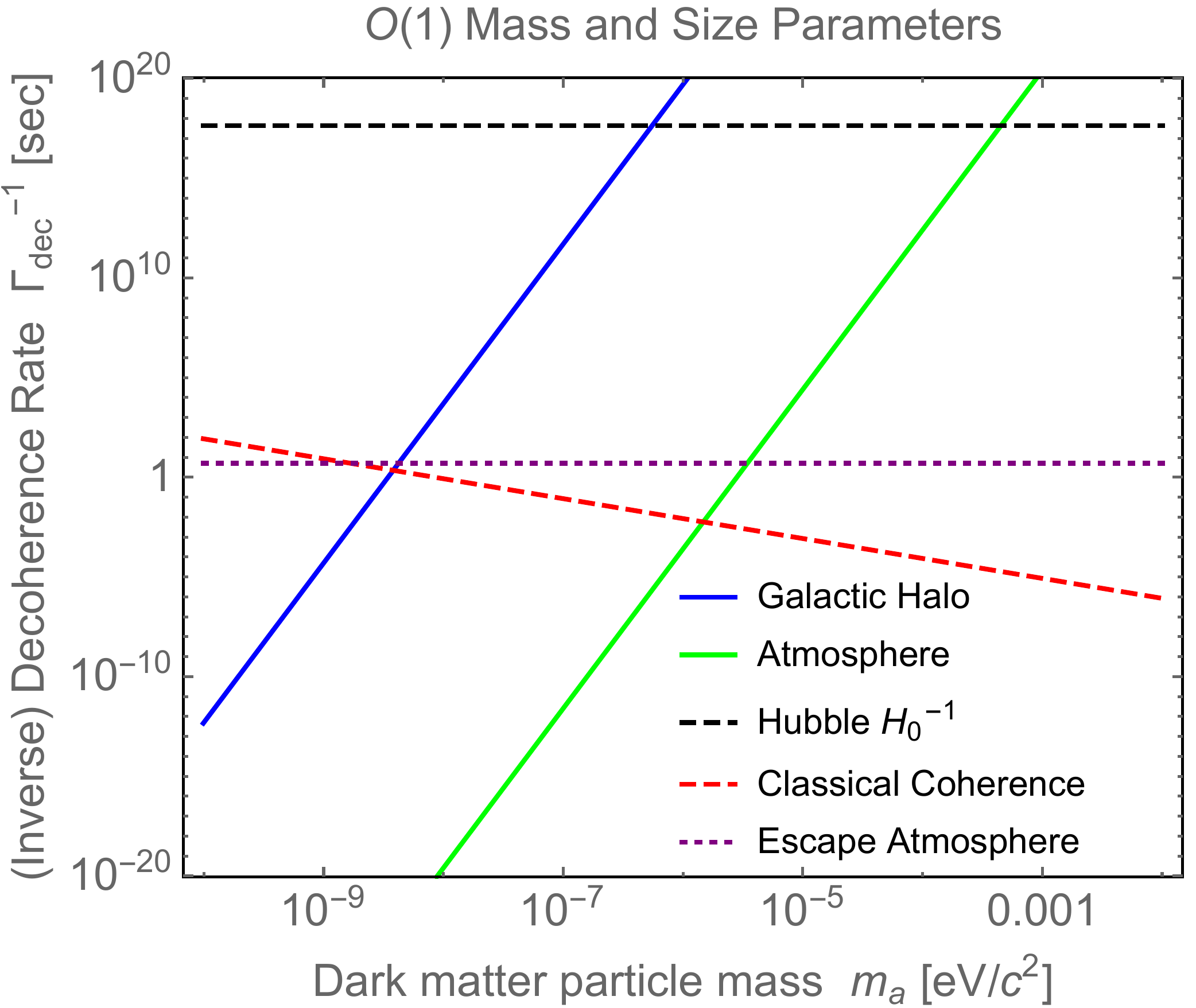}
\caption{The (inverse) decoherence rate is given in seconds for a dark-matter-Schr\"{o}dinger-cat-state (DMSCS) as a function of DM particle mass. (The case of the Gaussian density distribution is plotted, but it differs from the Yukawa result by an $\mathcal{O}(1)$ amount.) The mass distribution in the superposition is case (I) with $\beta_1=3/4$ and $\beta_2=5/4$. However, if we choose case (II) with $\beta_1=\beta_2=\mathcal{O}(1)$ and $\alpha_{1,2}$ different $\mathcal{O}(1)$ numbers, or case (III) with $\beta_1=\beta_2=\mathcal{O}(1)$, $\alpha_1=\alpha_2=\mathcal{O}(1)$ and $L\gtrsim1/\mu$, the results are similar.
Solid blue curve is for the galactic halo: The probe particles are taken to be protons (hydrogen) with density taken to be the typical matter density in the halo. Solid green curve is for the earth's atmosphere: The probe particles are $N_2$ molecules with density taken to be the density of air near the Earth's surface ($\sim1$\,kg/m$^3$). The black dashed line shows the current Hubble rate $H_0$, the red dashed line shows the coherence rate of the classical axion field $m_a v_a^2/(2\pi\hbar)$, and the purple dotted line shows the rate at which a DMSCS would escape the atmosphere.}
\label{fig:tdma}
\end{figure}

Putting all this together, if we take (I) $\beta_1/\beta_2=\mathcal{O}(1)$, or (II) $\alpha_1/\alpha_2=\mathcal{O}(1)$, or (III) $L\gtrsim1/\mu$, so that in each case there is an $\mathcal{O}(1)$ difference in the mass distribution of the DM states, we finally obtain a decoherence rate of
\begin{equation}
\Gamd \sim 10^{-20}\, \text{sec}^{-1}\left(\frac{10^{-6}\,\mbox{eV}}{m_a\,c^2}\right)^{\!8}\beta^2\alpha^2  \,\,\,\,\,\,\,(\mbox{galactic halo})
\end{equation}
This result is clearly very sensitive to the mass of the axion as it scales with $m_a^{-8}$. So depending on the choice of mass, the lifetime of the superposition state can be microscopic or it can be relevant on astrophysically long timescales. We find that to obtain a decoherence rate that is slower than the current Hubble rate $H_0$ (dashed black line), one would need to assume the axion mass to be $m_a\gtrsim 5\times10^{-7}$\,eV$/c^2\,(\beta\alpha)^{1/4}$, which is in fact quite reasonable for the QCD axion. 
Figure~\ref{fig:tdma} shows the relationship between the axion mass and the (inverse) decoherence rate as the solid blue curve, while the current Hubble rate $H_0$ is the dashed black curve.

For relevance to experimental searches for axions on earth, the DM sees an environment of much higher density provided by the atmosphere of the earth. For this we know the density is
\beq
\rho_p\approx1\,\mbox{kg/m}^3
\eeq
Here the relevant probe particles are atmospheric molecules, which are mainly nitrogen $N_2$. This enhancement in density and mass of the environment leads to much faster decoherence, and is given by the solid green curve in Figure~\ref{fig:tdma}. The corresponding decoherence rate is
\begin{equation}
\Gamd \sim 10^{3}\,\text{sec}^{-1}\left(\frac{10^{-6}\,\mbox{eV}}{m_a\,c^2}\right)^{\!8} \beta^2\alpha^2\,\,\,\,\,\,\,(\mbox{atmosphere})
\label{RateAtmosphere}\end{equation}
One of the relevant timescales to compare to for earth based experiments is the axion's own classical field coherence time. This arises due to the fluctuations in the axion's frequency over time $\Delta\omega \sim m_a\,v_a^2/\hbar$ due to the fact that it has a virialized spread of velocities in the galaxy. This is given by
\beq
\tcoh\sim {2\pi\hbar\over m_a v_a^2}
\eeq
In Figure~\ref{fig:tdma} this is indicated by the dashed red curve. For the quantum decoherence rate in the atmosphere to be slower than the classical coherence time, we find that $m_a\gtrsim 10^{-6}$\,eV$/c^2\,(\beta\alpha)^{2/9}$.

\begin{figure}[t]
\centering
\includegraphics[width=0.48\columnwidth]{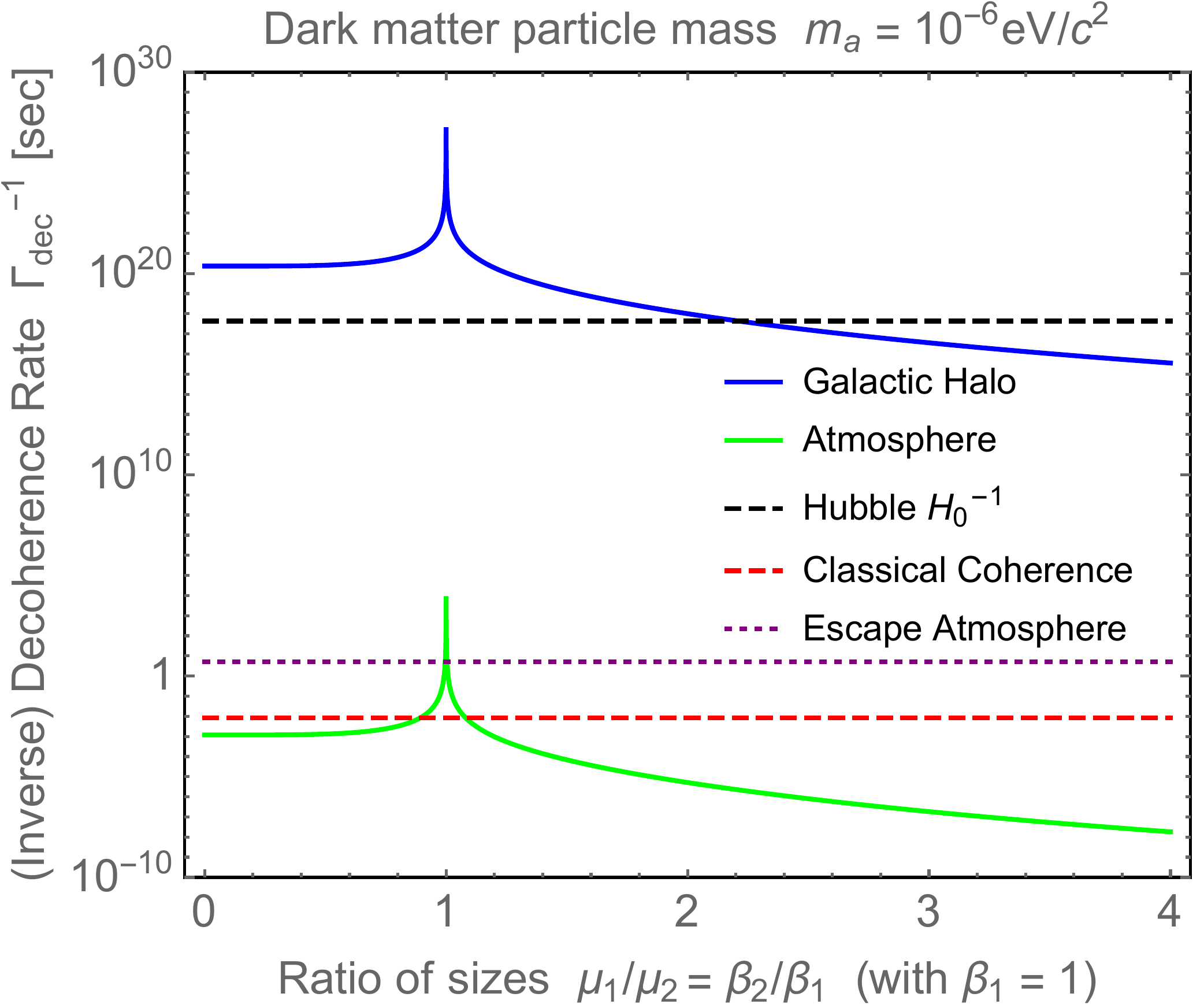}\,\,\,\,
\includegraphics[width=0.484\columnwidth]{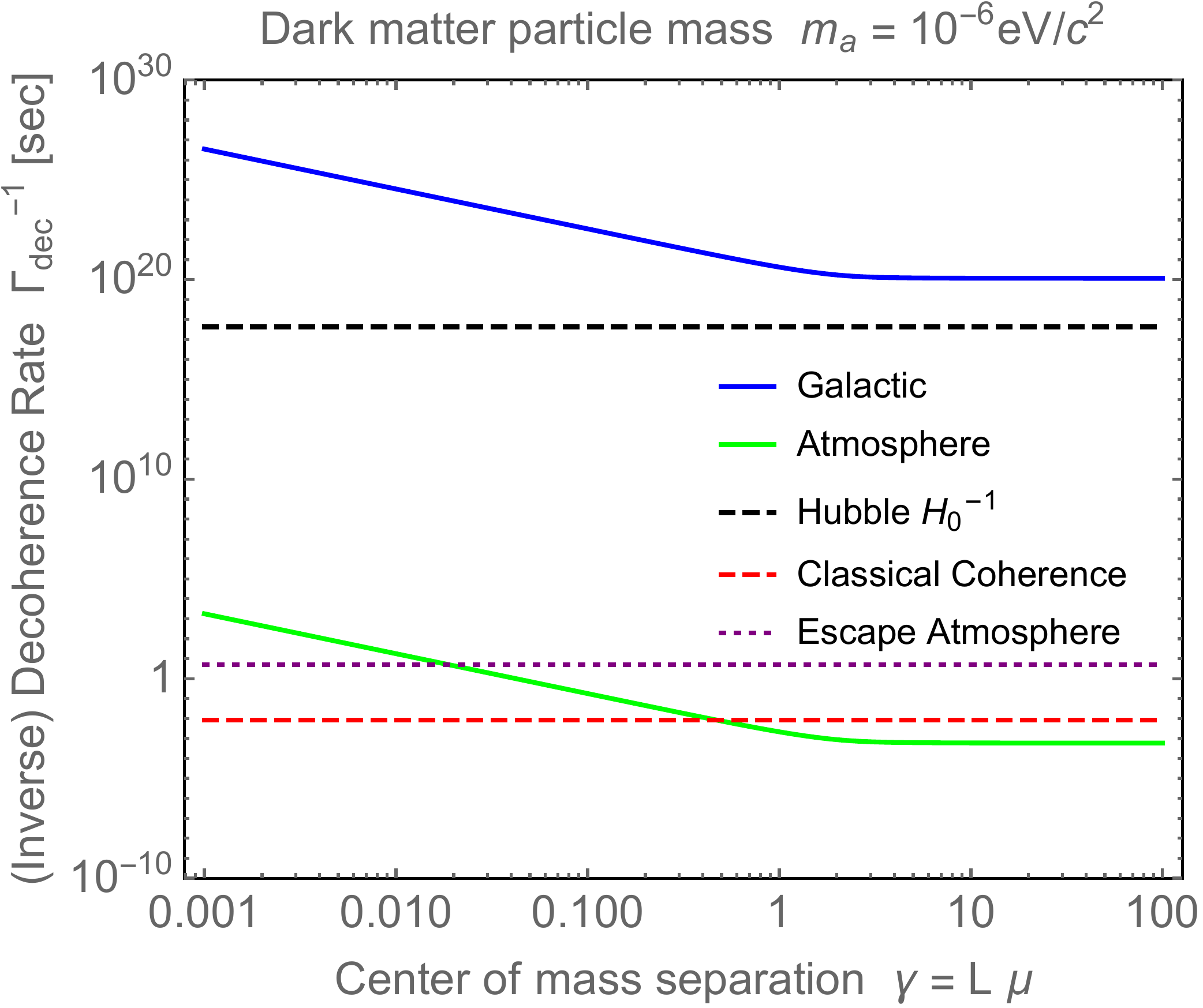}
\caption{The (inverse) decoherence rate is given in seconds for a dark-matter-Schr\"{o}dinger-cat-state (DMSCS) as a function of properties of the state. 
Left: Case (I) Horizontal axis is size and mass parameter $\beta_2=1/(\lambda_{dB}\mu_2)$, with $\beta_1=1/(\lambda_{dB}\mu_1)=1$, $\alpha_i=\beta_i^3$, and the same center of mass $L=0$. 
Right: Case (III) Horizontal axis is center of mass separation $\gamma=L\,\mu$ with $\beta_1=\beta_2=\alpha_1=\alpha_2=1$.
The meaning of the curves are the same as in Figure~\ref{fig:tdma}. The DM particles (axions) are taken to have a mass of $m_a=10^{-6}$\,eV$/c^2$; for other values of $m_a$, one simply re-scales the decoherence rates by $(10^{-6}\,\mbox{eV}/(m_a\,c^2))^8$ and re-scales the classical coherence rate by $(m_a\,c^2/(10^{-6}\,\mbox{eV}))$.  
Left: it can be seen that the choice of $\beta_2$ has a significant impact on the decoherence rate. A large difference between $\beta_1$ and $\beta_2$ corresponds to a superposition of very different macroscopic states, which decoheres more quickly, while a small difference corresponds to a superposition of very similar states, which decoheres more slowly $\Gamd\propto(\beta_1-\beta_2)^2$; see Eq.~(\ref{Xapprox}). 
Right: For small separations, there is naturally a slow rate of decoherence $\Gamd\propto \gamma^2$; see Eq.~(\ref{XapproxIII}). For large separations $L\gg1/\mu$ the rate approaches a constant.}
\label{fig:beta}
\end{figure}

We note that in case (I) the decoherence rate depends on the choice of $\beta_i$ quite sensitively. The closer the ratio of the $\beta_i$ is to 1, the smaller is the decoherence rate. Setting $\beta_1=1$, the (inverse) decoherence rate is plotted against $\beta_2$ in Figure~\ref{fig:beta} (left), which shows that the decoherence rate can be made quite small or large with different values of $\beta_2$. A value of $\beta_2$ near $1$ represents a superposition state which is very subtle, and therefore decoheres slowly (see Eq.~(\ref{Xapprox})), while a value of $\beta_2$ that differs significantly from $1$ is a superposition that is more noticeable and thus decoheres more quickly.

Similar behavior occurs in case (II) as we take the masses closer to each other, with $\potl\propto(\sqrt{\alpha_1/\alpha_2}-\sqrt{\alpha_2/\alpha_1})^2$, but we do not plot that here for the sake of brevity. Moreover, we note that in case (III) the decoherence depends sensitively on the separation $\gamma=L\,\mu$. This is seen in Figure~\ref{fig:beta} (right). For close separations of otherwise identically shaped objects, the decoherence is slow (see Eq.~(\ref{XapproxIII})), while at large separations it approaches a constant value, which is the usual scattering rate $\Gamd\to n\,\sigma\,v$. 

\subsection{Time Dependence}

Note that a DMSCS is only expected to quickly pass through the atmosphere as it moves through the galaxy. So the enhanced rate of Eq.~(\ref{RateAtmosphere}) will only apply briefly. For example, a DMSCS traveling at a relative speed of $220$\,km/sec for 1000\,km through the atmosphere takes only $\sim 5$\,sec. Hence one also needs $\Gamd\gtrsim 0.2$\,sec$^{-1}$ for the rate in the atmosphere to be relevant. This is given by the purple dotted line in the figures. Also note that for a DMSCS that passes through the earth (not just tangentially through the atmosphere), then the earth itself can act as the probe. However since its mass is so huge, we cannot ignore the back-reaction and the above analysis is not directly applicable. 

An important issue is that there can be significant spreading in the size of the overdensity. It is well known that a Gaussian wave packet,  that has an initial size $\mu_0^{-1}$,  spreads according to
\beq
\mu^{-2}(t) = \mu_0^{-2}+{\hbar^2 t^2 \mu_0^2\over m_a^2}
\label{muspread}\eeq
We note that while this is often applied to individual particle wave packets, it applies equally well to the axion field, which also obeys the non-relativistic Schr\"{o}dinger equation to first approximation. 
Since the decoherence rate $\Gamd\propto \mu^{-2}$ (recall Eq.~(\ref{Gammagravity})), the rate grows over time. If the initial decoherence rate, when $\mu=\mu_0$, is $\Gamma_{\tiny{\mbox{dec}},0}$, then the decoherence time $\tid$ (from Eq.~(\ref{tdecdef}) is the solution to
\beq
\tid + {\hbar^2\mu_0^4\over3 m_a^2}\,\tid^3 =1/\Gamma_{\tiny{\mbox{dec}},0}
\label{tspread}\eeq

\begin{figure}[t]
\centering
\includegraphics[width=0.6\columnwidth]{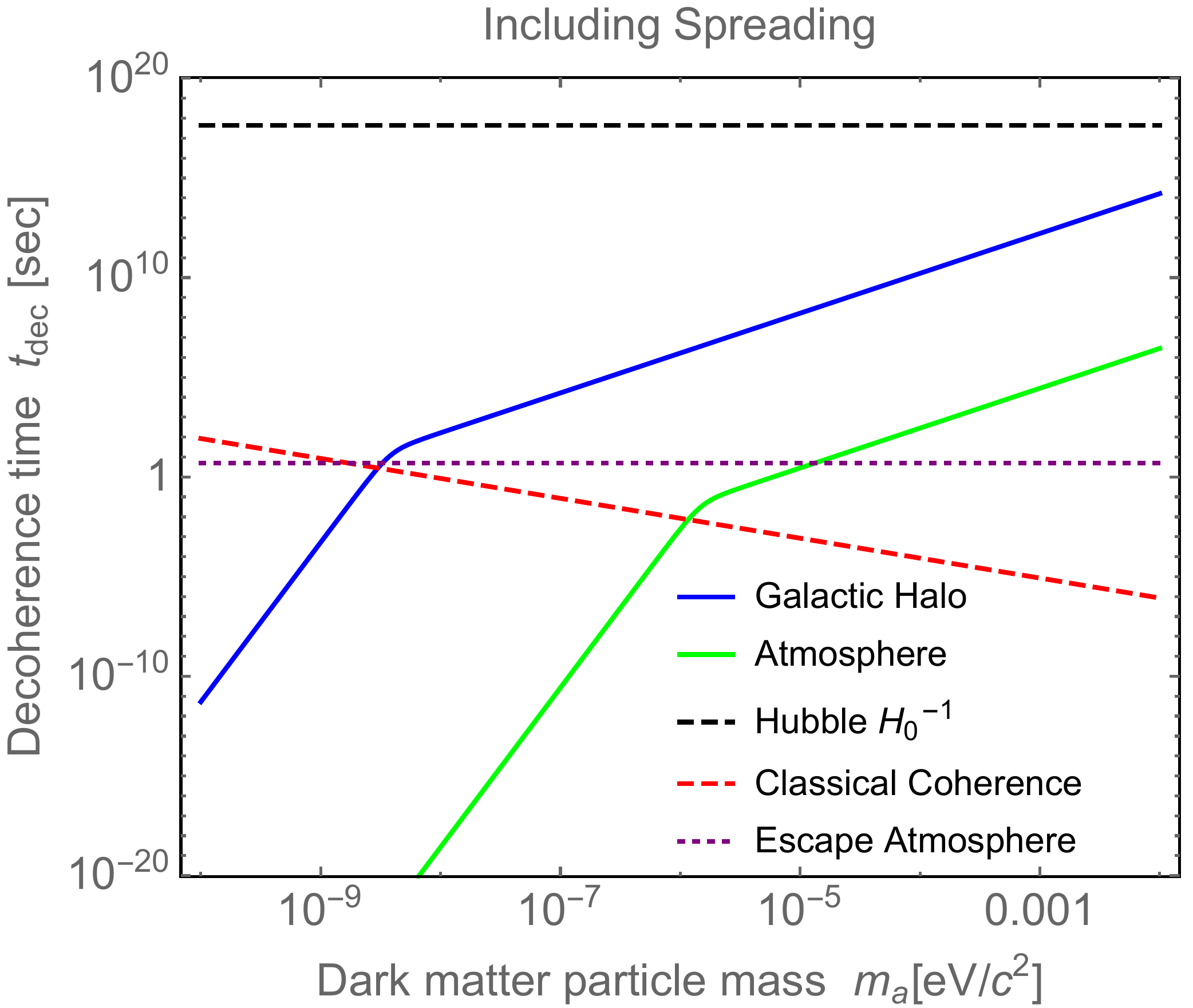}
\caption{The decoherence time is given in seconds for a dark-matter-Schr\"{o}dinger-cat-state (DMSCS) as a function of DM particle mass, where we take into account the effect of spreading of the size of the state over time. This is for case (II) with $\alpha_1=4/3$ and $\alpha_2=2/3$, $\gamma=0$, and initial value for widths $\beta_{10}=\beta_{20}=1$, i.e., the initial size is $\mu_0^{-1}=\lambda_{dB}$. We fix the masses, but allow the size to spread according to Eq.~(\ref{muspread}). The corresponding decoherence time is Eq.~(\ref{tspread}).}
\label{fig:spread}
\end{figure}

If we choose the initial value of $\mu_0$ to be $\mu_0\sim\lambda_{dB}$, as we did earlier, then the decoherence time in case (II) is plotted in Figure \ref{fig:spread}. For sufficiently small axion masses, the decoherence time is simply $\tid\approx1/\Gamma_{\tiny{\mbox{dec}},0}$, since there is very little time for spreading. However for larger axion masses, the decoherence time is $\tid\approx(3m_a^2/(\hbar^2\mu_0^4\,\Gamma_{\tiny{\mbox{dec}},0}))^{1/3}$, which shortens the decoherence time appreciably. 
For a DMSCS that is passing through the earth's atmosphere, there is a relatively small window in masses in which this spreading is important, beyond which it leaves the atmosphere too quickly. However, in the galactic halo, there is no such upper limit. Hence a spreading configuration will always decohere within the current age of the universe (unless we go to very high axion masses).   

Note that this spreading effect is not always as we described above. In particular, consider case (III), where the superposition is decribed by some separation between the center of masses $L$. Recall that the relevant dimensionless parameter is $\gamma=L\,\mu$. So as 
$\mu^{-1}$ increases in time due to the spreading, then the parameter $\gamma=L\,\mu$ may change too. If we consider the case in which the separation $L$ is independent of time, then $\gamma$ will decrease in time. Since $X\propto \gamma^2=(L\,\mu)^2$ for small $\gamma$ (ignoring possible logarithmic corrections), this leads to a rate that becomes essentially time independent. This leads to a decoherence time that is roughly estimated as $\tid\approx1/\Gamd$ and is therefore similar to the results of Fig.~\ref{fig:tdma}.

\subsection{Boson Stars}

Let us now comment on other interesting macroscopic structures; namely a Bose-Einstein condensate (BEC). These can emerge when large numbers of DM light scalar particles condense under the right conditions. Interestingly, some authors suggested that the condensate is extremely quantum mechanical and cannot be properly described using classical field theory \cite{Sikivie:2009qn,Erken:2011dz}. 

More specifically, the BEC turns out to be a gravitationally bound object of condensed scalars (axions) that arise from balancing the inward gravity against the kinetic pressure of a scalar field. These are sometimes called ``boson stars", see Refs.~\cite{Tkachev:1986tr,Gleiser:1988rq,Seidel:1990jh,Tkachev:1991ka,Jetzer:1991jr,Kolb:1993zz,Liddle:1993ha,Sharma:2008sc,Chavanis:2011zi,Chavanis:2011zm,Liebling:2012fv,Visinelli:2017ooc,Schiappacasse:2017ham,Hertzberg:2018lmt,Hertzberg:2018zte,Levkov:2018kau,Hertzberg:2020dbk}. On the main branch of stable solutions, the mass of the star $M$ and its radius $R$ are inversely related as
\beq
M\sim {\hbar^2\over G\,m_a^2\, R}
\label{MRrel}\eeq
Suppose they too were to organize into a DMSCS. Such objects will invariably carry a significant monopole (which is the mass $M$) and so our analysis here is not directly applicable, as there would formally be a divergence in the forward scattering direction. Furthermore, there can be large gravitational bending so that even a single probe particle could decohere a superposition of such states. To be quantitative, we can use our estimates in this work as a type of lower bound on the actual decoherence rate of such objects. If we return to Eq.~(\ref{Gammagravity}), take $\mu\sim 1/R$, $\chi\sim 1$, use Eq.~(\ref{MRrel}), and using the galactic protons as a probe, we obtain the bound (dropping $\mathcal{O}(1)$ numbers here) 
\beq
\Gamd\gtrsim {\hbar^2 m_p\rho_p\over v_p m_a^4}\sim10^{21}\,\mbox{sec}^{-1}\left(1\,\mbox{eV}\over m_a\,c^2\right)^{\!4}
\eeq
(one also needs $R\gg 100\,\mbox{km}\,(1\,\mbox{eV}/m_a)^2$ for $\Delta\ll1$). Hence, unless the axion mass is $m_a\gg 1$\,eV$/c^2$, the decoherence rate for boson stars is very rapid. This ensures that they are indeed very classical objects, which reinforces earlier work, including some of our own \cite{Guth:2014hsa,Hertzberg:2016tal}.

\section{Discussion}\label{Conclusions}

In this paper, we have computed the rate of gravitational decoherence of DM in a quantum superposition. In particular, we considered the case of a superposition of two mass densities, and we have shown that such a Schr\"{o}dinger-cat-like state has the potential to survive on astronomically relevant timescales depending on parameters, especially the DM particle (axion) mass. The DMSCS was modeled by an overdensity in the background DM of the galaxy surrounded by an underdensity, such that the total perturbation has zero mass (no monopole). For a spherically symmetric configuration, this has no multipoles at all and therefore such states are not seen appreciably by distant particles. Since mass is conserved, this is well defined. For non-spherical harmonics, one anticipates the rate of decoherence being faster; this decoherence can leave behind the spherical component in this superposed state, which we have focused on in this work. 

We have shown that for a range of typical axion masses, a stochastic perturbation of DM in the galaxy may indeed be in a coherent quantum superposition for an extended period of time. Specifically, for a superposition of overdensities whose mass distributions differ by an $\mathcal{O}(1)$ amount (either size, mass, or location of center of mass), and whose size and mass is set by the typical galactic de Broglie wavelength and density, we found that an axion particle mass of $m_a\approx 5\times10^{-7}$\,eV$/c^2$ corresponds to a decoherence rate of axions in the galactic halo of the current Hubble rate $H_0$. 
Moreover, we found that near the surface of the earth, an axion mass of $m_a\approx 10^{-6}$\,eV$/c^2$ corresponds to a decoherence rate comparable to the classical field's coherence rate, which is often a relevant timescale in ground-based experiments. Interestingly, in standard axion models, very light axion masses that are within a couple of orders of magnitude of these values are typical benchmark values as DM candidates. 

Furthermore, when spreading of the DMSCS is included, the rate of interactions can increase in time. This occurs, unless the only difference in the states is a fixed separation in center of masses. We found that with spreading, larger axion masses have a significantly faster decoherence rate, indicating that decoherence would eventually take place on a time-scale shorter than the current age of the universe. We also applied our results to DM BECs, which are boson stars, finding rapid decoherence and firmly establishing that they can be treated using classical field theory. 

We have focused our attention on only the leading order correction to the wave function of the probe particle, provided by the superposition of DM states. However, the probe particles can scatter off one another and other particles, which we have not taken into account. In the galactic halo, this scattering is very rare, as the mean free time is known to be extremely long, so we do not believe this will be important. However, near the surface of the earth, the probe particles (molecules) are rapidly scattering with one another. This may alter the decoherence rate;  we leave this for further investigation. Also, we have focussed here on ordinary matter causing decoherence of the axion DM quantum states, but in principle the diffuse DM axions can do it themselves. To define this, one could trace out, say, high momentum axions, and study the reduced density matrix of the remaining low momentum axions. 

Moreover, we have focused our attention purely on the ability of gravity to decohere the state. In the specific case of otherwise decoupled axions, we saw this can lead to long decoherence times with relatively natural choices for the axion mass. For the particular case of the QCD axion, one should also explore whether its non-gravitational interactions may also play a role in leading to decoherence. Microscopic models \cite{Kim:1979if,Shifman:1979if,Dine:1981rt,Zhitnitsky:1980tq,Kim:2008hd} usually involve dimension 5 operators that couple the axion to gauge bosons, $\sim g_{a\gamma\gamma}\phi F_{\mu\nu}\tilde{F}^{\mu\nu}$, as well as to fermions $\sim g_{aff}\partial_\mu\phi\bar{\psi}\gamma^\mu\gamma^5\psi$. These interactions may also lead to decoherence and deserve investigation. 

Though we have shown that the superposition state of the DM may survive for observationally interesting timescales, one may wonder about the possibility for such a state to exist in the first place. To estimate the natural rate of formation for the superposition state, one can determine a rate of the evolution of the fluctuations in the axion field; related work includes  Refs.~\cite{Sikivie:2009qn,Erken:2011dz,Guth:2014hsa,Levkov:2018kau,Dvali:2017ruz}. By considering the evolution of the axion field in the non-relativistic limit, an interaction rate can be estimated. It would be interesting if interactions can lead to appreciable spreading of wave functions, as is often the case due to chaos. By making simple estimates for the spreading rate for the configurations discussed in the previous sections, our decoherence rate results indicate that the spreading of the axion wave function is slower than its decoherence, rendering the state always classical. 
We note that if instead the Schr\"{o}dinger-cat state comes about from some other, quicker dynamical process, then the timescale of decoherence may be considered relatively large, and such a state may exist for a long time.
In the future, the mechanism and likelihood of the formation of such Schr\"{o}dinger-cat states should be explored in detail. 

Our analysis may be useful in other interesting contexts. For example, in non standard models of dark matter, such as the “superfluid dark matter” scenario \cite{Berezhiani:2015bqa} or ultralight axions, one is often essentially treating the dark matter in the framework of classical field theory. So in some sense, decoherence is implicitly assumed. For very large dark matter configurations, such as relevant to an entire galaxy or its core, the decoherence will be rather rapid, indicating that this classical assumption should be justified. Furthermore, the phenomenon of gravitation leading to decoherence may have relevance to understanding the transition to classicality of quantum fluctuations from inflation. 

Further, the observational signatures and implications of these states have yet to be considered. Since some properties of Schr\"{o}dinger-cat states at high occupancy can often be modeled using classical ensemble averaging \cite{Hertzberg:2016tal}, any direct observational consequences are far from clear. One may need to construct and measure some novel correlation functions of non-commuting operators. We leave all these issues for future consideration.

\section*{Acknowledgments}
We would like to thank Mark Gonzalez, Andi Gray, Mudit Jain, Jacob Litterer, Fabrizio Rompineve, Neil Shah, and Shao-Jiang Wang for discussion. This research was supported in part by the Munich Institute for Astro- and Particle Physics (MIAPP) which is funded by the Deutsche Forschungsgemeinschaft (DFG, German Research Foundation) under Germany's Excellence Strategy – EXC-2094 – 390783311.  MPH is supported in part by National Science Foundation grant PHY-1720332.

\appendix
\section{Supplementary Material}\label{app:overlapcomp}

\subsection{Scattered Wave Function}\label{app:scf}

The full time dependence of the scattered wave function at large radii, assuming a very wide wave packet, is provided by the following function
\ba
\varphi_s(r,t)\amp\equiv\amp \frac{1}{4 d^2 k} \Bigg[\left(2 d^2 k+i r\right) e^{\frac{m \left(2 d^2 k+i r\right)^2}{4 d^2 m+2 i \hbar t}} \Big(\text{Erf}\left(\frac{2 d^2 k+i r}{\sqrt{4 d^2+\frac{2 i \hbar t}{m}}}\right)+1\Big)\nonumber\\
\amp\amp\,\,\,\,\,\,\,\,\,\,\,\,\,+\left(2 d^2 k-i r\right) e^{\frac{m \left(2 d^2 k-i r\right)^2}{4 d^2 m+2 i \hbar t}} \text{Erfc}\left(\frac{2 d^2 k-i r}{\sqrt{4 d^2+\frac{2 i \hbar t}{m}}}\right)\Bigg]\label{varphi} 
\ea
where ``Erf'' and ``Erfc'' denote the error function and conjugate error function respectively.

\subsection{Gaussian Profile}\label{app:gauss}

Here we provide some additional details of the calculation of the wave function overlap for a Gaussian overdensity followed by underdensity. The density profile is given below, along with the resulting Newtonian potential, Fourier transform, and scattering amplitude.

\ba
\delta\rho_G(r)\amp=\amp M \mu^3\zeta_G(\mu r),\,\,\,\,\,\mbox{with}\,\,\,\,\,\zeta_G(\hat{x})= \!\left(1-\frac{2}{3}\hat{x}^2\right) e^{-\hat{x}^2}\label{deltarhogapp}\\
\Phi_{N,G}(r)\amp=\amp-\frac{2\pi}{3}GM\mu e^{-\mu^2 r^2}\\
\hat{\zeta}_G(\dv)\amp=\amp\frac{\pi^{3/2}\dv^2}{6}e^{-\frac{\dv^2}{4}}\\
f_G(k,\theta)\amp=\amp f^{(1)}_{G}({\bf k}',{\bf k}) =\frac{\pi^{3/2}}{3}\frac{G M m^2}{\hbar^2\mu^2}e^{-\frac{\dv^2}{4}}\label{fg}
\ea
Now, following the integrals of Section \ref{overlapints}, we can compute the individual parts of the overlap of the wave functions that scatter off of this Gaussian potential. By focusing on the physically important case of $k\gg \mu$, we find
\beq
\ch_{ij}=\frac{\pi^3 \mu_i \mu_j}{18(\mu_i^2+\mu_j^2)}
\eeq

\subsection{Yukawa Profile}\label{app:yuk}

Here we repeat the details of the above appendix, but for the case of a density profile that gives rise to a Yukawa potential. The details are as follows
\ba
\delta\rho_Y(r)\amp=\amp M \mu^3\zeta_Y(\mu r),\,\,\,\,\,\mbox{with}\,\,\,\,\,\zeta_Y(\hat{x})=\left(\delta^3({\bf\hat{x}})-\frac{1}{4\pi \hat{x}}\right)e^{-\hat{x}}\label{deltarhogapp}\\
\Phi_{N,Y}(r)\amp=\amp-\frac{GM}{r}e^{-\mu r}\\
\hat{\zeta}_Y(\dv)\amp=\amp\frac{\dv^2}{1+\dv^2}\\
f_Y(k,\theta)\amp=\amp f^{(1)}_{Y}({\bf k}',{\bf k}) =\frac{2GMm^2}{\hbar^2\mu^2}\frac{1}{\dv^2+1}\label{fy}
\ea
Computing the integrals of Section \ref{overlapints}, and taking once more $k\gg\mu$, we find
\beq
\ch_{ij}=\frac{\mu _i \mu _j \log \left(\frac{\mu _i}{\mu _j}\right)}{\mu _i^2-\mu _j^2}
\eeq

\end{document}